\documentclass[journal=jacsat,manuscript=article]{achemso}

\usepackage[version=3]{mhchem} 
\usepackage{graphicx}  
\usepackage{caption} 
\usepackage{subcaption} 
\usepackage{balance}
\usepackage{mathptmx}
\usepackage[format=plain,justification=justified,singlelinecheck=false,font={stretch=1.125,small,sf},labelfont=bf,labelsep=space]{caption}
\usepackage{float}
\usepackage{fancyhdr}
\usepackage{fnpos}
\usepackage{array}
\usepackage{droidsans}
\usepackage{charter}
\usepackage[T1]{fontenc}
\usepackage[usenames,dvipsnames]{xcolor}
\usepackage{setspace}
\usepackage[compact]{titlesec}
\usepackage{hyperref}

\author{Kanad Sen}
\email{kanadsen01@gmail.com}
\affiliation[IIT Bombay]
{Department of Mechanical Engineering, IIT Bombay, Mumbai, India}
\author{Saksham Gupta}
\email{drsakshamgupta@gmail.com}
\affiliation[CMINDS, IIT Bombay]
{Centre for Machine Intelligence and Data Science, IIT Bombay, Mumbai, India}
\author{Abhishek Raj}
\affiliation[IIT Bombay]
{Department of Mechanical Engineering, IIT Bombay, Mumbai, India}
\author{Alankar Alankar}
\affiliation[IIT Bombay]
{Department of Mechanical Engineering, IIT Bombay, Mumbai, India}

\title[An \textsf{achemso} demo]
  {Graph Residual based Method for Molecular Property
Prediction}


\begin{document}

\begin{tocentry}
    \centering
    \includegraphics[width=1\textwidth]{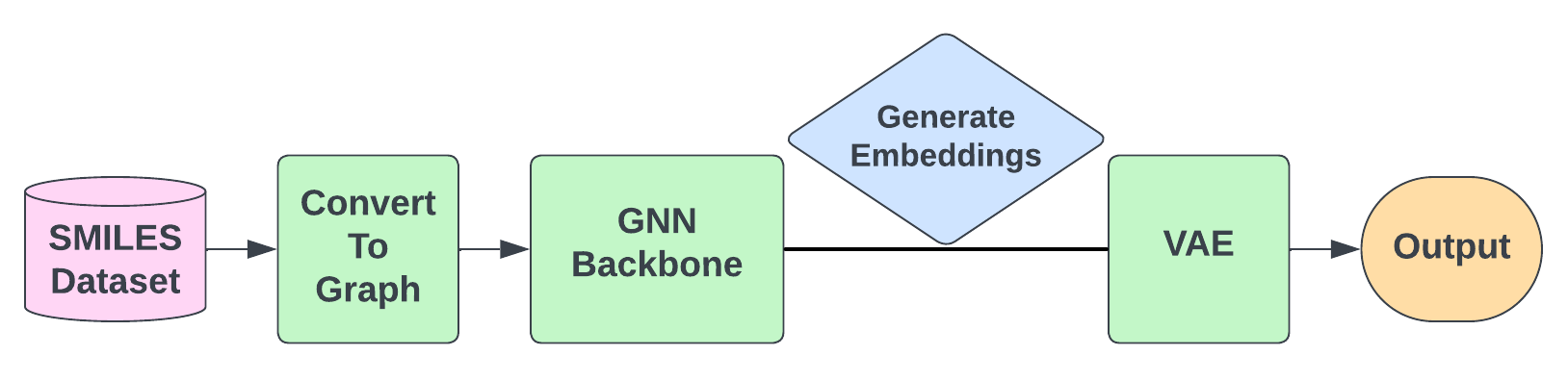} 
    \newline
    \textbf{Overall Pipeline for Molecular Property Prediction} 
\end{tocentry}

\begin{abstract}
\noindent
Machine learning-driven methods for property prediction have been of deep interest. However, much work remains to be done to improve the generalization ability, accuracy, and inference time for critical applications. The traditional machine learning models predict properties based on the features extracted from the molecules, which are often not easily available. In this work, a novel Deep Learning method, the Edge Conditioned Residual Graph Neural Network (ECRGNN), has been applied, allowing us to predict properties directly only the Graph-based structures of the molecules. SMILES (Simplified Molecular Input Line Entry System) representation of the molecules has been used in the present study as input data format, which has been further converted into a graph database, which constitutes the training data. This manuscript highlights a detailed description of the novel GRU-based methodology, ECRGNN, to map the inputs that have been used. Emphasis is placed on highlighting both the regressive property and the classification efficacy of the same. A detailed description of the Variational Autoencoder (VAE) and the end-to-end learning method used for multi-class multi-label property prediction has been provided as well. The results have been compared with standard benchmark datasets as well as some newly developed datasets. All performance metrics that have been used have been clearly defined, and their reason for choice.
\end{abstract}

\section{Introduction}\label{intro}
\noindent Quantitative structure-property relationship (QSPR) a traditional method developed as early as 1935 \cite{QSPR}, has provided an efficient framework to facilitate computation on chemical properties by breaking them down into their smallest blocks. The mathematical model allows the molecules to be broken down according to their chemical, structural, physical, and biological properties. The method of property prediction consists of two steps, which involve breaking down the molecule into property descriptors and then regressing on these descriptors. The molecular descriptors generate a \textit{n}-dimensional data with  \begin{math}d=[d_1,d_2, \ldots, d_n]^T\end{math} which indicates the presence of a particular feature as generated by the molecular descriptor. 

Molecular descriptors can be 0D, 1D, 2D, as well as 3D. \cite {Cit_3, Cit_4} The most common type of descriptors is 0D descriptors, typically used in conjunction with Machine Learning models. \cite{Cit_33} The 0D molecular descriptors are simply atoms and molecular properties that are used to describe the molecule. \cite{Cit_9} However, these descriptors do not contain any information about the relationship between the atoms. 1D descriptors can be considered functional sub-groups or a molecule's building blocks. It is the most commonly used descriptor, yet these don’t contain any structural information. Thus, most of the work that has been done using this method to this day lacks accuracy. In more recent times, the use of 2D descriptors has been used, which allows the molecular topology to be represented in its true form and introduces several geometry-based features that can be represented in graph format, like cyclicity, symmetry, and connectivity between atoms. Lastly, a more modern higher dimensional descriptors \cite{Cit_21}, which can represent complex molecular geometries and dynamics, has been used \cite{Cit_18, Cit_19, Cit_20}; however, exploiting such features is computationally expensive. 

QSPR modelling is challenging as it requires subject expertise to identify the functional subgroups. Automating the process requires a careful selection of algorithms and data to identify the information descriptor that is to be used in the Machine Learning model. This inherently poses a significant challenge since such descriptors cannot be generalized for all tasks. Thus, alternate ways have come up to extract molecular embeddings or fingerprints directly from the molecular structure rather than from its molecular descriptors. This is where Graph Neural Networks (GNNs) offer efficient methodology. A molecule can be arranged in a graph-like structure from which deep learning modules directly extract relevant features. This kind of learning on non-Euclidean datasets is also known as geometric learning and has been made more easily possible due to the recent development of the Pytorch-geometric module. \cite{Cit_22}

\section{Graphical Representation of Molecule }\label{graph_repre_mol}

GNNs are recognized as a potent tool for extracting knowledge from graph-structured data, showcasing impressive performance across a variety of domains, like social networks, molecular structures, drug screening, chip design, traffic flows, recommendation systems, and knowledge graphs. In various tasks such as node classification, graph classification, graph regression, and link prediction, GNNs consistently achieve state-of-the-art performance. The concept of GNN was first introduced by Gori et al. \cite{Cit_10}, where they extended the recursive neural networks to give rise to a new set of networks that is more node-focused. This work was further extended by Scarselli et al. \cite{Cit_11}, which created an end-to-end GNN network along with modern backpropagation techniques. The authors provided a detailed mathematical description of the GNN method in this work. One of the major important properties of graphs and GNNs is that they are size-agnostic, i.e., the sample size can differ from one input to another, and they are permutation invariant. This property provides a major advantage to GNNs over CNN-based architectures since the input graph size and the number of nodes and edges can easily vary, whereas, in CNNs, we must provide a fixed set of input sizes.

There are various variety of tasks that can be performed by GNNs, such as node level tasks \cite{Cit_41, Cit_42}, edge level tasks \cite{Cit_43}, and Graph level tasks \cite{Cit_44, Cit_45}. The problem of Molecule property prediction, which we are trying to solve, is a Graph-level task where each molecular graph generates an embedding, which is to be used further for prediction. The most primitive GNN used for solving such types of problems is the Message Passing Neural Network (MPNN). \cite{Cit_49} MPNN is based on the simple Graph Convolution format, which allows each node to learn from its subsequent neighbouring nodes, and that information is then aggregated into the original node. Similar to the Graph Convolutions that are used in MPNN, several other aggregators have been proposed, most notable being Graph Attention Network (GAT) based on the multi-head attention mechanism \cite{Cit_39}, the recently developed GraphSAGE (Graph Sample and Aggregated) \cite{Cit_38}, a framework for inductive representation learning on large graphs which samples and aggregates feature from a node's neighbourhood to learn its embedding, the GraphConv mechanism \cite{Cit_37}  etc. Graph Isomorphism Network (GIN) by Xu et al. \cite{Cit_40}  provides flexibility in capturing the local and global graph structure, making it suitable for regression tasks.

Jiying et al. \cite{Cit_31}  presented an approach to fine-tune Graph Neural Networks by framing the problem within an Optimal Transport framework, capitalizing on graph topology. By introducing a masked Optimal Transport problem and defining the Masked Wasserstein distance, the Graph Induced Optimal Transport (GTOT) -Tuning framework facilitates efficient knowledge transfer from pre-trained models while preserving local feature invariances. GTOT leverages graph structure to enhance knowledge transfer and offer flexibility for integration with complementary techniques like contrastive learning. \cite{Cit_51} However, this method poses a high computational demand and has limited efficacy on small datasets or those with sparse graph structures, as is usually the case with medicinal and molecular data.

Another enhancement on GNNs is the Graph Multiset Transformer (GMT), introduced by Jinheon et al. \cite{Cit_32}  as a graph multi-head attention-based global pooling layer. It captures interactions and relations between nodes of a graph based on its structural dependencies. This ensures that GMT satisfies both injectiveness and permutation invariance. While this leads the model to learn more accurate representations of entire graphs by capturing structural dependencies between nodes, training a transformer model such as GMT requires significant computational resources. The inner workings of GMT also capture complex structural dependencies, which are hard to interpret in a physical manner. 

Multilabel classification is widely explored in the image modality, with extensive use of transformer-like encoder-decoder models, as shown by Shichao et al. \cite{Cit_24}  and Shilong et al. \cite{Cit_25}. Graphical classification also relies heavily on transformer-based approaches, with GNN-based models severely lacking performance. Various methods exist for multi-class multi-label methods on different modalities of input data. \cite{Cit_52, Cit_53} In this context, we are trying to explore the classification capabilities of GNNs. The first paper that proposed this was a simple Graph Convolutional Network (GCN) structure \cite{Cit_46}, which was used for multi-label classification. 

With the introduction of the "Attention is all you need" paper \cite{Cit_23} and recent advances in Transformer architecture, Shermukhamedov et al. \cite{Cit_26} introduced ElemBert – v1 and ElemBert – v2, which employ transformer mechanisms for binary classification utilizing structural information. These models, known for their flexibility, demonstrate adaptability across various datasets. Through benchmarking, they have showcased the state-of-the-art performance of Elembert in diverse material property prediction tasks, encompassing both organic and inorganic compounds. Additionally, Heyrati et al. \cite{Cit_27}  proposed BioAct-Het, leveraging a heterogeneous Siamese neural network to capture the intricate relationship between drugs and bioactivity classes, thereby consolidating them within a unified latent space. While this approach enhances the identification of drug discovery markers, it faces limitations in effectively handling the simultaneous classification of multiple labels. Notably, both approaches require separate models for training each label, resulting in heightened training and inference times and increased resource requirements per data point.

Alpersein et al. \cite{Cit_28} introduce a new method called the All SMILES Variational Autoencoder (VAE), focusing on enhancing the optimization of molecular properties through VAEs that work with both SMILES string and graph-based molecular representations. This approach handles encoding a molecule's multiple SMILES strings through a layered arrangement of recurrent neural networks. \cite{Cit_28} It aggregates the hidden states corresponding to each atom across these representations, employing an attention mechanism to consolidate them into a comprehensive, fixed-size latent vector. The model's ability to decode this representation back into a distinct set of SMILES strings for the original molecule allows it to establish a nearly one-to-one relationship between the molecule and its latent representation. This effectively addresses the issue of SMILES strings' non-uniqueness and sidesteps the computational demands associated with graph convolution operations. VAEs are particularly advantageous in this context because they are adept at learning intertwined representations of complex data distributions. Additionally, VAEs incorporate a regularization mechanism by enforcing a prior distribution on the latent space, which aids in mitigating overfitting and enhances the model's ability to generalize. 

This work emphasizes the prediction of both the regression and classification properties of molecules. It presents results for new datasets and compares them with well-established benchmarks. The implemented Graph-based classification shows significance in the realm of molecular property prediction, where molecular structures are represented as graphs, with atoms as nodes and chemical bonds as edges, playing a crucial role in predicting various molecular properties such as toxicity, solubility, bioactivity, and drug efficacy, among others. \cite{Cit_27, Cit_35} Intricate structural features and relationships inherent in molecular compounds can be captured by leveraging graph-based representations, enabling more accurate and comprehensive predictions. Tasks within molecular property prediction often include binary classification to distinguish between toxic and non-toxic compounds, multiclass classification to categorize compounds into different therapeutic classes or activity levels, and regression to predict quantitative properties like molecular weight or solubility. These tasks are pivotal in drug discovery, materials science, and environmental health.

This paper attempts to introduce a novel architecture, a Gated Recurrent Unit (GRU) - based residual edge convolutional network to capture the inherent pattern in the molecular graphs and introduce long-range dependencies. A complete model overview is presented along with its mathematical formulation to explain the underlying architecture. Apart from that, a diagrammatic representation of the model in an end-to-end fashion has been presented to better understand the workflow for both regression and classification tasks. This study's highlight, along with its explainability in relation to the underlying chemistry. Additionally, a comparison has been made between the present SOTA model and the performance of benchmark datasets. This study has also described a detailed statistical analysis of the classification datasets and data processing.

\subsection*{SMILES Encoding}

Morgan's technique \cite{Cit_13} for generating unique machine descriptions and the establishment of the CAS ONLINE search system were significant developments for the description of the chemical structure database. The widespread adoption of computers in the field of chemical nomenclature led to the prevalent use of line notation. Instead of drawing detailed molecular structures, line notation uses lines, symbols, and characters to convey essential structural information. This is attributed to the computational efficiency with which computers can handle linear data strings. \cite{Cit_15} Rapid advancements in computer technology have outpaced the incremental growth of chemical information. The synergy between computer capabilities and chemical knowledge allows for the storage of all existing chemical information on current hardware. The historical challenge of merely storing information in machines has been largely overcome. Current and future efforts are now focused on creating highly efficient systems that deliver chemically relevant information when required. To achieve this, a new chemical language and associated computer programs are being developed, grounded in the innovative information system known as SMILES. \cite{Cit_14} 

The SMILES system was created to be genuinely interactive with computers. Its ease of use relies on computer programs that meticulously reformat the input from the chemical user. This achievement stemmed from the pursuit of specific original objectives, including the unique description of a chemical structure graph. This system was designed to overcome difficulties and provide a user-friendly experience by leveraging computer interactivity and stringent input-recoding processes. 

\begin{figure}[hbt!]
\centering
\captionsetup{justification=centerlast} 
\includegraphics[width=0.75\textwidth]{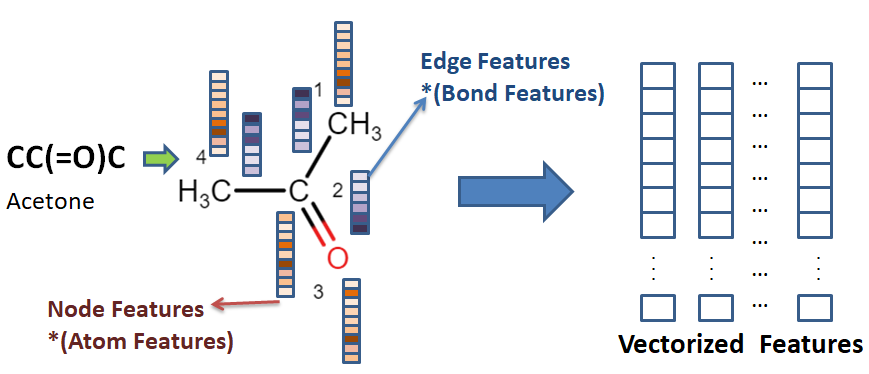}
\caption{Schematic representation of preprocessing of SMILES data.}
\label{fig:Fig1}
\end{figure}

SMILES is a line notation used to encode molecular structures, allowing for a linear representation of the molecule so that a computer program can understand the structure. It specifically represents a valence model of a molecule and is widely used in chemistry to represent molecular structures in a compact and human-readable form. The graph of a chemical structure was to be uniquely described, including but not limited to the molecular graph comprising nodes (atoms) and edges (bonds). \cite{Cit_16} Since most of the datasets come in SMILES \cite{Cit_14} encoded format, they need to be converted into a graph database for use in GNNs. A molecular graph can be represented in the form of $G(n)= (V, E)$ where $n$ is the number of nodes or vertices, i.e the atoms in the molecule and $e_{v,w}$ represents the edges i.e the atomic bonds in the molecule where $e_{v,w} \epsilon E$. A typical example of the molecule in this format has been shown in Figure \ref{fig:Fig1}.

Several types of atomic and bond features are present; however, for this study, only the basic features available without data-specific feature engineering are chosen to allow for the model's generalizability. Based on Figure \ref{fig:Fig1}, a sample encoding structure of the features is presented in Tables \ref{tab:Table_1} and \ref{tab:Table_2} along with their encodings. These numerical encodings are then converted to one-hot encoded vectors for use in the geometric dataset. The tables show a small sample of the node and edge feature set.
\begin{table}[hbt!]
\centering
\captionsetup{justification=centerlast}
\caption{Representation of encoding of edge features}
\label{tab:Table_2}
\small
\begin{tabular}{lccccc}
\hline
\textbf{Edge} & \textbf{Bond} & \textbf{Conjugated} & \textbf{Aromatic} & \textbf{Is In Ring} & \textbf{Connects} \\
\hline
Edge 1 & Single & No & No & No & (1,2) \\
Edge 2 & Double & No & No & No & (2,3) \\
Edge 3 & Single & No & No & No & (2,4) \\
\hline
\end{tabular}
\end{table}
\begin{table}[hbt!]
\centering
\captionsetup{justification=centerlast}
\caption{Representation of encoding of node features}
\label{tab:Table_1}
\small
\begin{tabular}{lccccc}
\hline
\textbf{Node} & \textbf{Atom Type} & \textbf{Atomic Number} & \textbf{Neighbour} & \textbf{Hydrogen} & \textbf{Formal Charge} \\
\hline
Node 1 & C & 6 & 1 & 3 & 0 \\
Node 2 & C & 6 & 3 & 0 & 0 \\
Node 3 & O & 8 & 1 & 0 & 0 \\
Node 4 & C & 6 & 1 & 3 & 0 \\
\hline
\end{tabular}
\end{table}

A complete representation of the number of features used for both the node and edge cases has been shown in Tables \ref{tab:Table_4} and \ref{tab:Table_3}. They also represent the size of the feature space of the one-hot encoded vectors. Proper dimension size and the type of features with detailed descriptions are presented.

\begin{table}[hbt!]
\centering
\captionsetup{justification=centerlast}
\caption{Edge feature encoding descriptions}
\label{tab:Table_4}
\small
\resizebox{\textwidth}{!}{%
\begin{tabular}{llcc}
\hline
\textbf{Edge Features} & \textbf{Description/Exemplary Values} & \textbf{Number of Features} & \textbf{Encodings} \\
\hline
Bond type & Single, Double, Triple and Aromatic & 4 & [0--3] \\
Conjugated & Whether the bond is conjugated & 1 & [4] \\
Is in ring & Whether the bond forms an aromatic ring & 1 & [5] \\
\hline
\end{tabular}%
}
\end{table}

\begin{table}[hbt!]
\centering
\captionsetup{justification=centerlast}
\caption{Node feature encoding descriptions}
\label{tab:Table_3}
\small
\resizebox{\textwidth}{!}{%
\begin{tabular}{llcc}
\hline
\textbf{Node Features} & \textbf{Description/Exemplary Values} & \textbf{Number of features} & \textbf{Encodings} \\
\hline
Atom type & All possible atoms in the periodic table & 117 & [0--116] \\
Bonds & Number of bonds the atom is involved in & 9 & [117--125] \\
Charge & Formal charge of the atom & 1 & [126] \\
Hybridization & sp, sp\textsuperscript{2}, sp\textsuperscript{3}, sp\textsuperscript{3}d, sp\textsuperscript{3}d\textsuperscript{2}, sp\textsuperscript{2}d, s, other, unspecified & 9 & [127--135] \\
Is aromatic & Indicates if the atom forms any aromatic system & 1 & [136] \\
Is in ring & Indicates if the atom forms any ring & 1 & [137] \\
Atomic number & Atomic number & 1 & [138] \\
Hs & Number of bonded hydrogen atoms & 9 & [139--147] \\
\hline
\end{tabular}%
}
\end{table}

\section{Model Description}\label{sec:methods}

\subsection{Edge Conditioned Residual Graph Neural Network (ECRGNN)}\label{GNN_Block}
\label{GNN_block}
A significant amount of work in the field of GNNs, \cite{Cit_62, Cit_63}  focuses on the utilization of node features to find different patterns and relations between different molecule structures but fails to consider the edge features. These are particularly important when talking about organic molecules since information such as the number of bonds and aromaticity influence various properties of organic molecules.

Thus, this paper considers Edge Conditioned Graph Convolution (ECC) Neural Network \cite{Cit_6}, which modifies the normal message passing algorithms of the traditional GNN to include a correction made by using the relevant edge features- aromaticity, multiple bond presence, resonance, etc. Using an Artificial Neural Network (ANN), representations of the edge features are obtained in the feature space of the node features, and then add them to the node features during message passing algorithm, ensuring that the resulting node features obtained have information not just from the surrounding neighbouring nodes, but from the edge features connecting those nodes as well.

\begin{equation} \label{1_eqn}
x_i' = \theta x_i + \sum_{j \in N(i)} x_j \cdot h_{\theta}(e_{ij})
\end{equation}

The Eq.\ref{1_eqn} describes the convolution operation of the ECC, where $h_{\theta}$ is a multi-layer perceptron which takes in the edge features $(e_{ij})$ and dynamically gives the required edge weight. Here, $ x $ represents the node features that are passed to the model, and $\theta$ represents the parameters of the model to be updated.

While this solves the inherent problems within implementations of ECRGNN, the training is optimized with several other methods. It has been observed that many molecules have similar substructures, such as aldehyde groups, benzene rings, and other common elements. Since these groups often contribute to properties such as solubility, boiling point, etc., in a standard manner across different molecules, a mechanism has been introduced to store these influences in a separate state so that this information can travel through various molecules. Bigger molecules will thus contribute information that can act as a template for smaller molecules, resulting in faster computation efficiency and more focused learning. Gated Graph Neural Network (GGNN) \cite{Cit_45} tries to store temporal states by induction of a GRU unit in place of each graph node of a GNN during the message passing layer. Since there is a need for a spatial log of the states that are observed by the algorithm, a GRU layer is applied each at the end of the “hops”, i.e. the end of the message passing algorithm, saving the state as needed to preserve node properties across different molecules. The GRU update steps at the $l^{th}$ layer are\\

Update gate:
\begin{equation*}
    z_i^{(l)} = \sigma(W_z^{(l)}x_i^{(l)} + U_z^{(l)}h_i^{(l-1)} + b_z^{(l)} ), \tag{2a}
\end{equation*}

Reset gate:
\begin{equation*}
    r_i^{(l)} = \sigma(W_r^{(l)}x_i^{(l)} + U_r^{(l)}h_i^{(l-1)} + b_r^{(l)} ), \tag{2b}
\end{equation*}

Candidate hidden state: 
\begin{equation*}
    \tilde{h}_i^{(l)} = \tanh(W_h^{(l)}x_i^{(l)} + U_h^{(l)}(r_i^{(l)} \odot h_i^{(l-1)}) + b_h^{(l)} ), \tag{2c}
\end{equation*}

Final hidden state update:
\begin{equation*}
    h_i^{(l)} = (1 - z_i^{(l)}) \odot h_i^{(l-1)} + z_i^{(l)} \odot \tilde{h}_i^{(l)}. \tag{2d}
\end{equation*}



This approach clearly delineates the flow of information from one layer to the next, emphasizing the sequential dependency between the ECC output and GRU input within the same layer and the GRU output serving as the ECC input for the subsequent layer.

\begin{figure}[hbt!]
    \centering
    \captionsetup{justification=centerlast} 
    \includegraphics[width=0.5\textwidth]{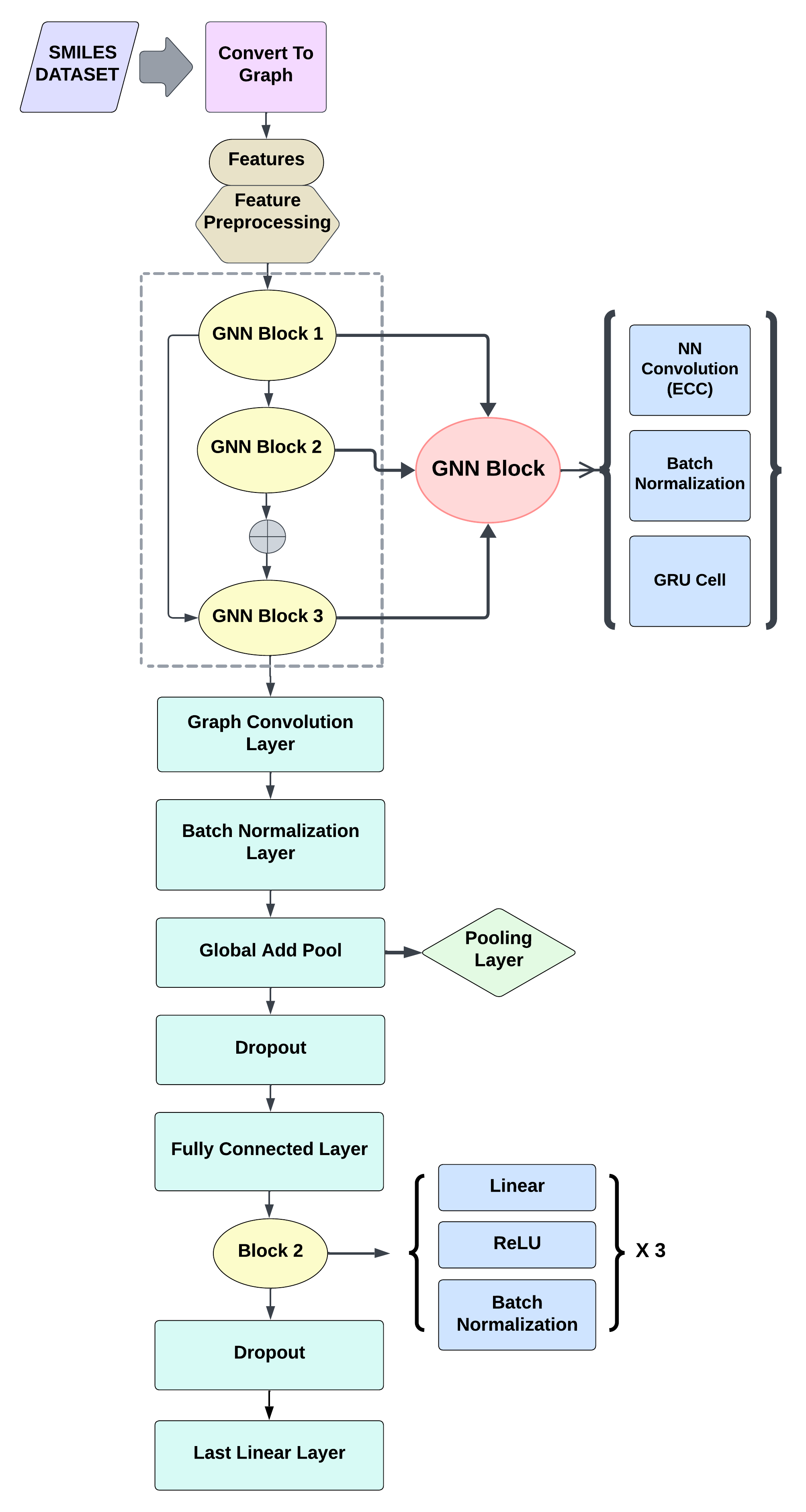}
    \caption{Schematic diagram of the flow of ECRGNN model.}\label{fig:GNN_fig}
\end{figure}

A pictorial representation of the ECRGNN model is shown in Figure \ref{fig:GNN_fig}, which describes in detail the various parts of the model. The figure shows a combined diagram of the ECRGNN along with the attached VAE model for classification purposes. As evident from Figure \ref{fig:GNN_fig} and \ref{fig:VAE_fig}, layers of GNN blocks are added, which consist of an ECC layer, Batch Normalization layer, and GRU cell. The number of layers added to the model represents the k-hops distance from the node from which the model aggregates information. In this case, since 3 layers are added, each node is considered to aggregate information 3 nodes away. This is especially useful in the case of larger molecules.

However, in this case, deeper models have also been experimented with. Experimentation has been done with upto 5 hop structure, but it did not lead to better accuracy of the results, rather the model performed worse. Kipf and Welling \cite{Cit_54} showed that increasing the number of hops could decrease the performance due to the inclusion of noise from the additional hop structures. Several methods have been proposed to solve this problem similar to Residual connections in Computer Vision, e.g., Highway GCN by Rashimi et al. \cite{Cit_55}, and Jump Knowledge Network by Xu et al. \cite{Cit_56}.

In this implementation, a simple Residual network connection is used, deriving inspiration from the DeepGCN model of Li et al. \cite{Cit_58}, who used Residual Connections from the Resnet model \cite{Cit_67}. A simple version of ResGCN \cite{Cit_57} is used in this case, whereas ECRGNN has a single residual block, aggregating information from the first and second hop, which is to be fed in the third hop. Since all molecular graphs are not large, deeper layers are not used. However, a future study will include a proper accurate comparison of Residual graph networks on molecular structure.

Additionally, a GraphConv Mechanism \cite{Cit_37} as described in Eq. \ref{3_eqn} is applied after the third hop, with pooling at the top of it. This transforms the node and edge features into a feature space where all the edge features are 1, thus filtering all the remaining information of the edge features into the node features. This is then passed through a pooling layer to obtain a representation of the molecule that can be passed through any subsequent ANN for classification or regression tasks. The implementation of the same is described by 
\begin{equation}\label{3_eqn}
x_i' = W_1 x_i + W_2 \sum_{j \in N(i)} e_{ij} x_j , \tag{3}
\end{equation}
where $W_1$ and $W_2$ are the parameters and $e_{ji}$ is the edge weight which by default is considered as 1. The $x$ here represents the node features computed from the successive ECC blocks.

The ANN is defined as a simple multilayer perceptron consisting of 3 hidden layers of 128 dimensions each and an output layer. Based on the Regression task, the output layer dimension is 1, and for the classification task, the output layer dimension is 256 since the embeddings generated by the model are passed to the VAE as described in Sec. \ref{exap_gan_algorithm}.

\subsection{Explanation of the GNN algorithm}
\label{exap_gan_algorithm}
\noindent In the current model, a GRU layer after each hop is added. The addition of the GRU layer, along with the residual connection, increases the model's performance. A logical explanation along with outputs of some explainers like GNNExplainer \cite{Cit_47} and PGExplainer \cite{Cit_61} has been presented as proofs of the working of the model.

The molecular graph inputs are dynamic in nature, i.e. their size varies. Due to this, the model encounters molecules of different lengths during training and testing. Despite training the model on a random selection of training set, the model may find it difficult to adjust the weights to a varying set of inputs. Thus, GRU layers are implemented after each hop, along with a residual connection between the first and the third hop, to solve this problem and allow better adaptability to the molecules.

To understand the detailed working and effectiveness of the GRU and residual connections, GNNExplainer \cite{Cit_47} has been used to present an explanation on the behaviour of the model for selected target property on two datasets, Lipophilicity \cite{Cit_34} and Aqueous solubility \cite{Cit_34}.\\






\begin{figure*}
    \captionsetup{justification=centerlast} 

    \centering
    \begin{subfigure}{0.4\textwidth}
        \captionsetup{justification=centerlast} 
        \includegraphics[width=0.9\linewidth, height=0.9\linewidth]{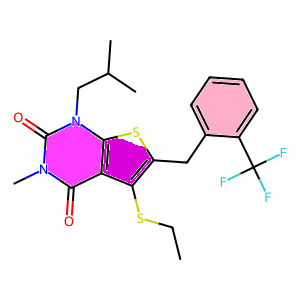}
        \caption{}
        \label{fig:7a}
    \end{subfigure}
    \begin{subfigure}{0.45\textwidth}
        \captionsetup{justification=centerlast} 
        \includegraphics[width=1\linewidth, height=0.7\linewidth]{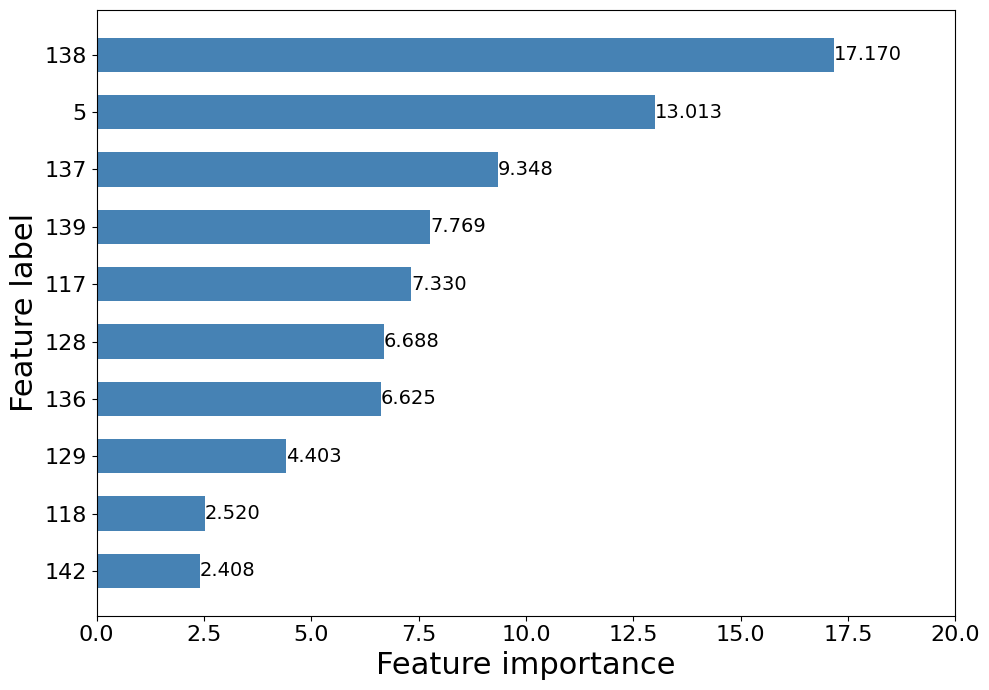}
        \caption{}
        \label{fig:7b}
    \end{subfigure}


    \begin{subfigure}{1\textwidth}
        \centering
        \includegraphics[width=1.0\linewidth]{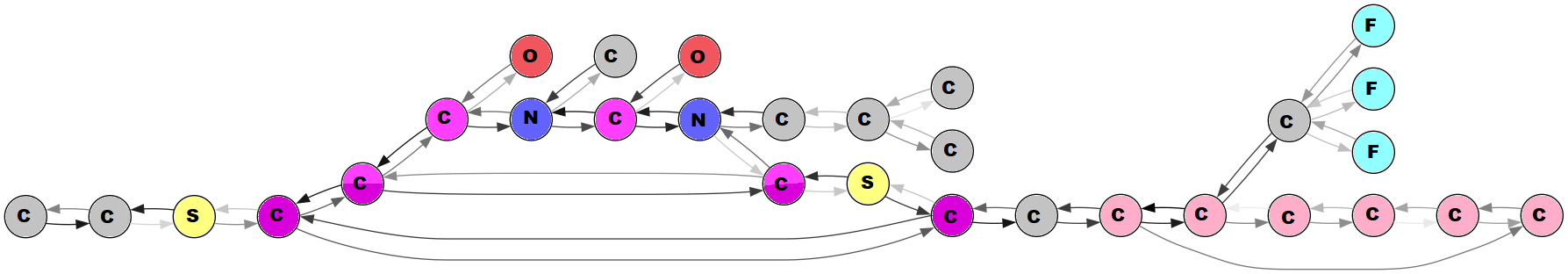}
        \captionsetup{justification=centerlast} 
        \caption{}
        \label{fig:7c}
    \end{subfigure}
    \begin{subfigure}{1\textwidth}
         \centering
        \includegraphics[width=1.0\linewidth]{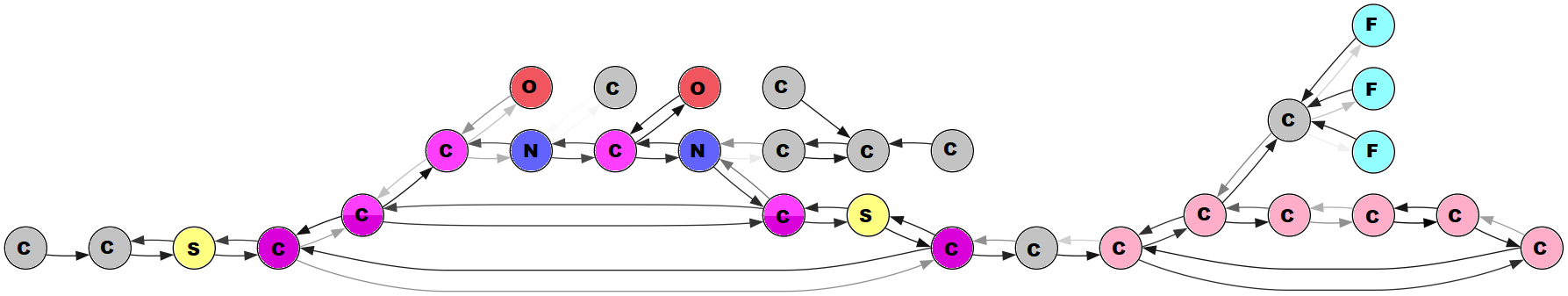}
        \captionsetup{justification=centerlast} 
        \caption{}
        \label{fig:7d}
    \end{subfigure}

    \caption {Comparative Representation of the outputs of GNNExplainer on the target Molecule Name: \textit{5-ethylsulfanyl-3-methyl-1-(2-methyl propyl)-6-[[2-(trifluoromethyl)phenyl]methyl] thieno [2,3-d]pyrimidine-2,4-dione}, (a) Molecular Structure, (b) Node Feature Importance,  (c) Output generated by the ECC model, (d) Output generated by the ECRGNN model. In this case, the target property considered is Lipophilicity. (The grayscale value of the arrows signifies the strength of the relationship between the two nodes, which directly correlates to the predicted property)}
    \label{fig:main}
    
\end{figure*}


\begin{figure}
    \captionsetup{justification=centerlast} 
    \centering

    \begin{subfigure}{0.4\textwidth}
         \centering
         \captionsetup{justification=centerlast} 
        \includegraphics[width=0.9\linewidth, height=0.9\linewidth]{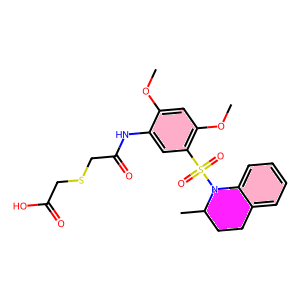}
        \caption{}
        \label{fig:8a}
    \end{subfigure}
    \begin{subfigure}{0.45\textwidth}
         \centering
         \captionsetup{justification=centerlast} 
        \includegraphics[width=1\linewidth, height=0.7\linewidth]{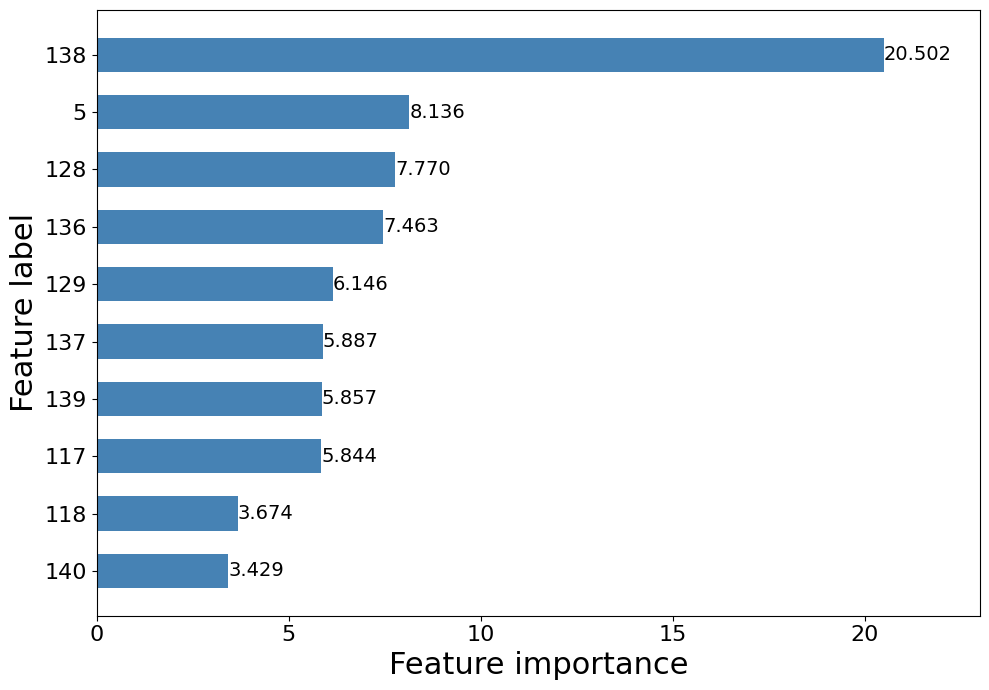}
        \caption{}
        \label{fig:8b}
    \end{subfigure}


    \begin{subfigure}{1\textwidth}
         \centering
         \captionsetup{justification=centerlast} 
        \includegraphics[width=1\linewidth]{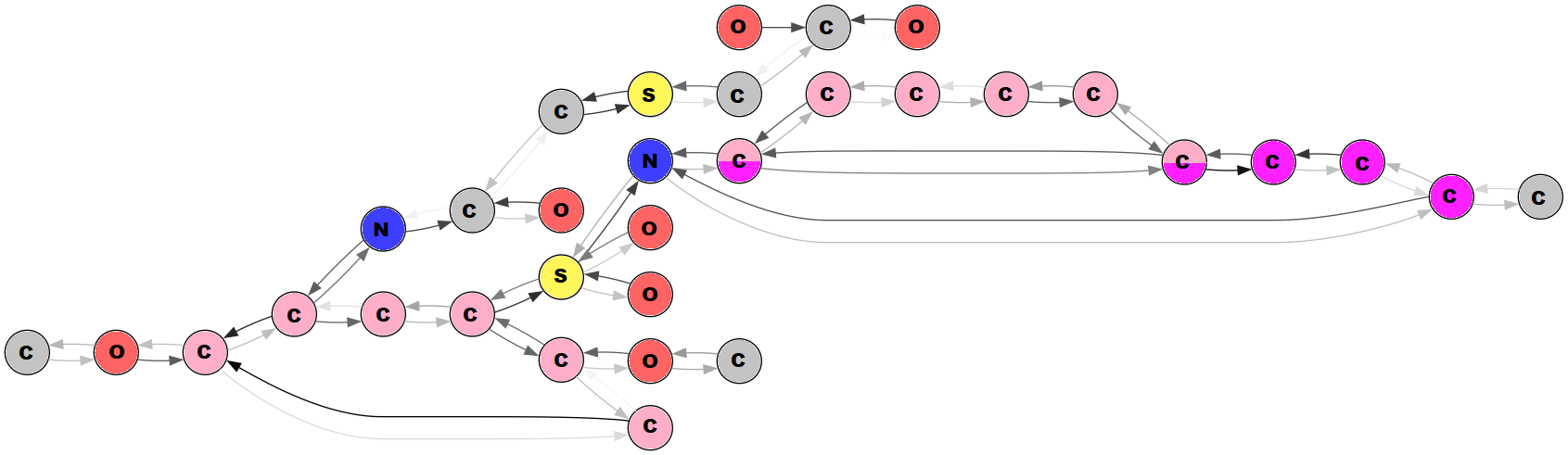}
        \caption{}
        \label{fig:8c}
    \end{subfigure}
    \begin{subfigure}{1\textwidth}
         \centering
         \captionsetup{justification=centerlast} 
        \includegraphics[width=1\linewidth]{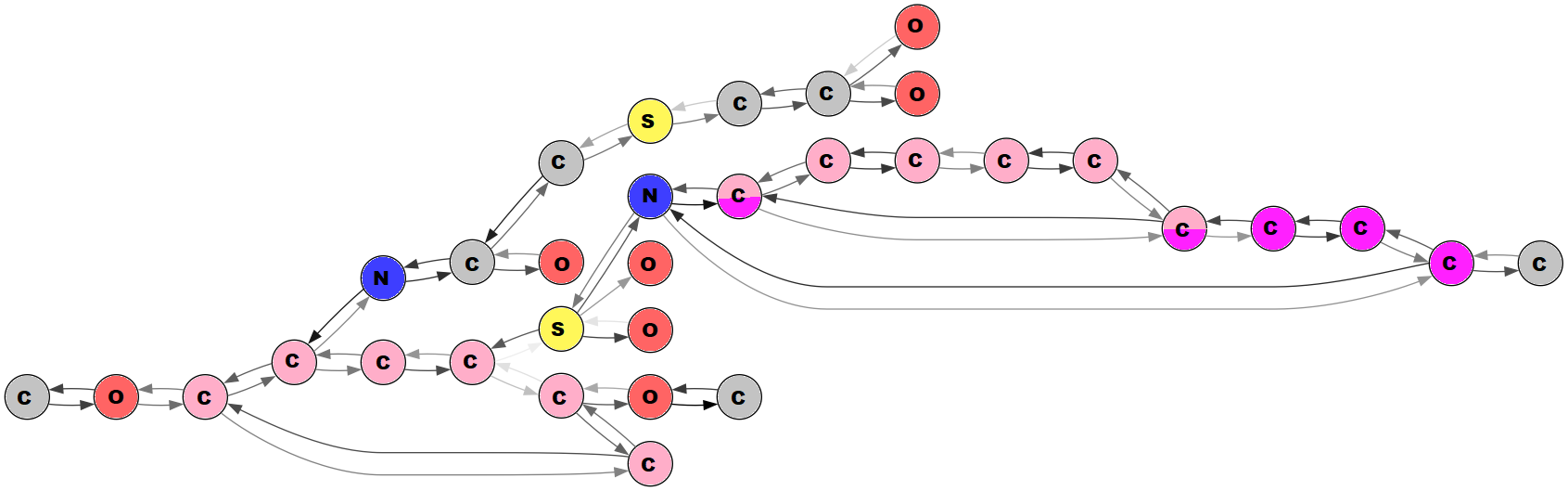}
        \caption{}
        \label{fig:8d}
    \end{subfigure}

    \caption{Comparative Representation of the outputs of GNNExplainer on the target Molecule Name: \textit{2-[2-[2,4-dimethoxy-5-[(2-methyl-3,4-dihydro-2H-quinolin-1-yl)sulfonyl]anilino]-2-oxoethyl]sulfanylacetic acid,} (a) Molecular Structure, (b) Node Feature Importance, (c) Output generated by the ECC model, (d) Output generated by the ECRGNN model. In this case, the target property considered is Lipophilicity. (The grayscale value of the arrows signifies the strength of the relationship between the two nodes, which directly correlates to the predicted property)}
    \label{fig:main}
\end{figure}


\begin{figure*}
    \captionsetup{justification=centerlast} 
    \centering
    \begin{subfigure}{0.4\textwidth}
        \captionsetup{justification=centerlast} 
        \includegraphics[width=0.9\linewidth,width=0.9\linewidth ]{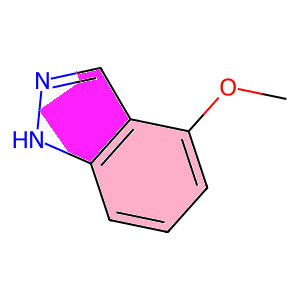}
        \caption{}
        \label{fig:3a}
    \end{subfigure}
    \begin{subfigure}{0.45\textwidth}
        \captionsetup{justification=centerlast} 
        \includegraphics[width=1\linewidth, height=0.7\linewidth]{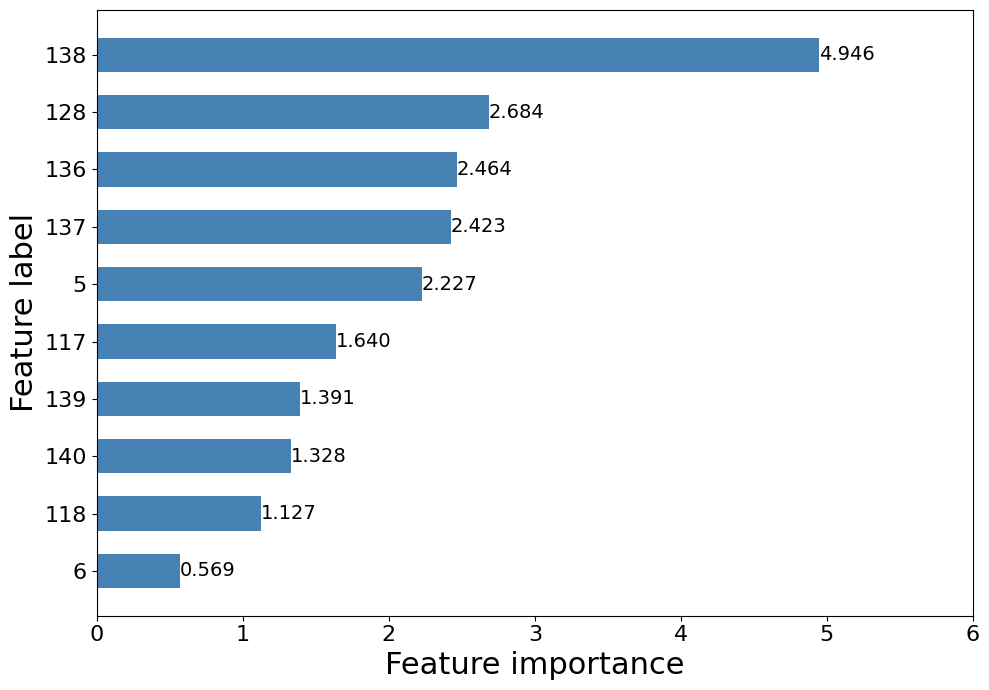}
        \caption{}
        \label{fig:3b}
    \end{subfigure}


    \begin{subfigure}{1\textwidth}
        \centering
        \captionsetup{justification=centerlast} 
        \includegraphics[width=1\linewidth]{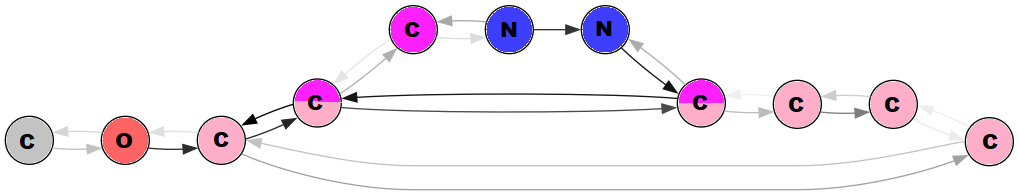}
        \caption{}
        \label{fig:3c}
    \end{subfigure}
    \begin{subfigure}{1\textwidth}
         \centering
         \captionsetup{justification=centerlast} 
        \includegraphics[width=1\linewidth]{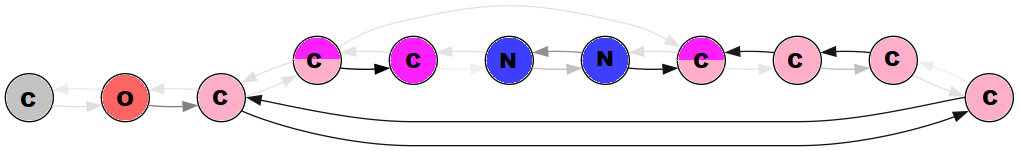}
        \caption{}
        \label{fig:3d}
    \end{subfigure}

    \caption{Comparative Representation of the outputs of GNNExplainer on the target Molecule Name: \textit{4-methoxy-1H-indazole (COc1cccc2[nH]ncc12)}, (a) Molecular Structure, (b) Node Feature Importance, (c) Output generated by the ECC model, (d) Output generated by the ECRGNN model. In this case, the target property considered is Lipophilicity. (The grayscale value of the arrows signifies the strength of the relationship between the two nodes, which directly correlates to the predicted property)}
    \label{fig:main}
\end{figure*}


\begin{figure*}
    \centering
    \captionsetup{justification=centerlast} 
    \begin{subfigure}{0.4\textwidth}
        \includegraphics[width=0.9\linewidth, height=0.9\linewidth]{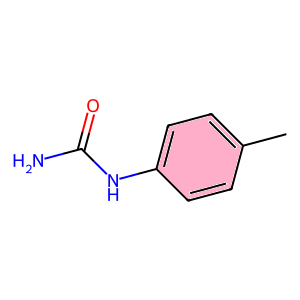}
        \caption{}
        \label{fig:4a}
    \end{subfigure}
    \begin{subfigure}{0.45\textwidth}
        \includegraphics[width=1\linewidth, height=0.7\linewidth]{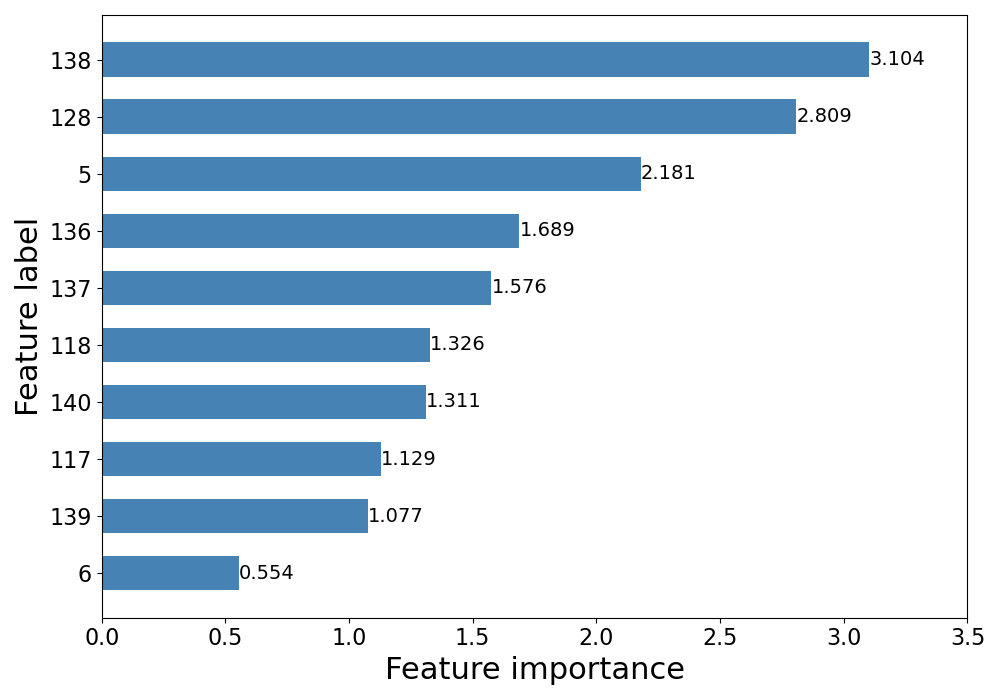}
        \caption{}
        \label{fig:4b}
    \end{subfigure}


    \begin{subfigure}{1\textwidth}
         \centering
        \includegraphics[width=0.8\linewidth]{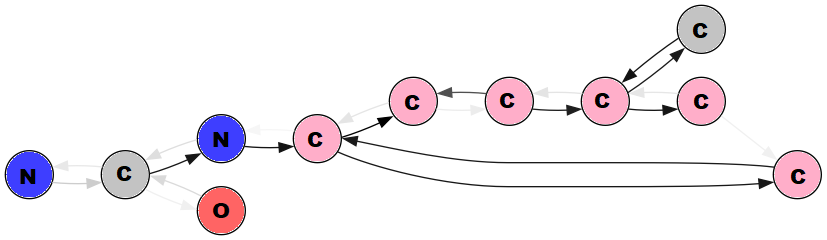}
        \caption{}
        \label{fig:4c}
    \end{subfigure}
    \begin{subfigure}{1\textwidth}
         \centering
        \includegraphics[width=0.8\linewidth]{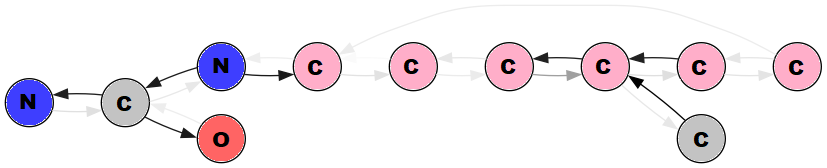}
        \caption{}
        \label{fig:4d}
    \end{subfigure}

    \caption{Comparative Representation of the outputs of GNNExplainer on the target Molecule Name: \textit{(4-methylphenyl)urea}, (a) Molecular Structure, (b) Node Feature Importance, ,(c) Output generated bt the ECC model, (d) Output generated by the ECRGNN model.  In this case, the target property considered is Aqueous Solubility. (The grayscale value of the arrows signifies the strength of the relationship between the two nodes, which directly correlates to the predicted property)}
    \label{fig:main}
\end{figure*}


\begin{figure*}
    \centering
    \captionsetup{justification=centerlast} 
    \begin{subfigure}{0.4\textwidth}
        \includegraphics[width=0.9\linewidth, height=0.9\linewidth]{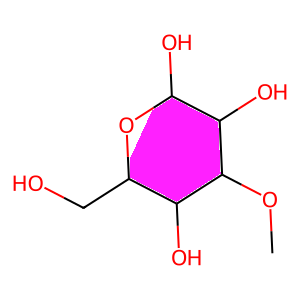}
        \caption{}
        \label{fig:5a}
    \end{subfigure}
    \begin{subfigure}{0.45\textwidth}
        \includegraphics[width=1\linewidth, height=0.7\linewidth]{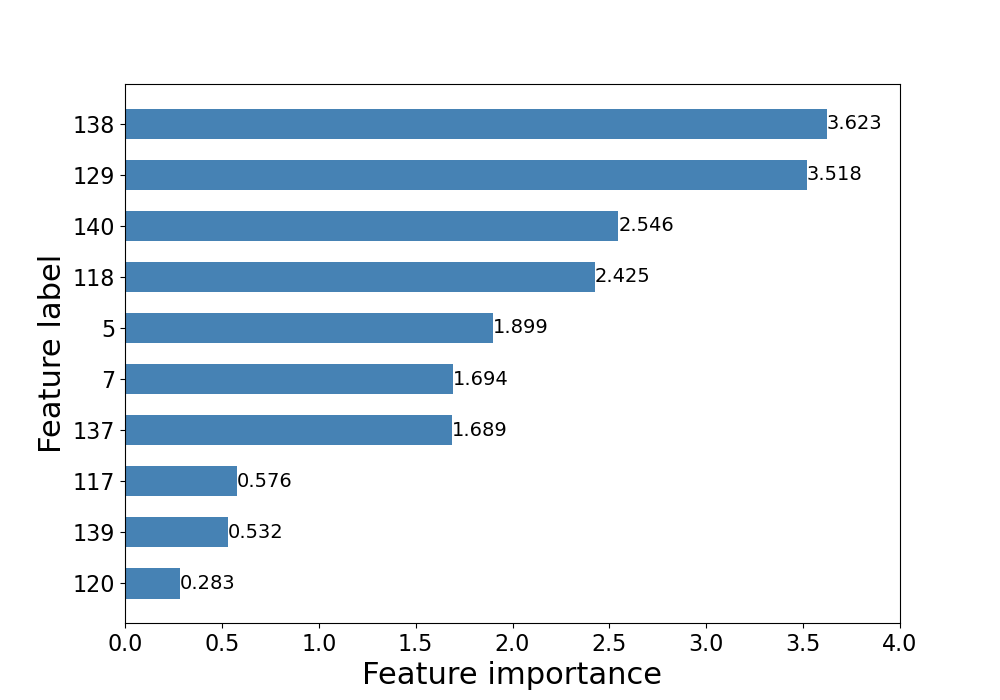}
        \caption{}
        \label{fig:5b}
    \end{subfigure}


    \begin{subfigure}{1\textwidth}
        \centering
        \includegraphics[width=0.8\linewidth]{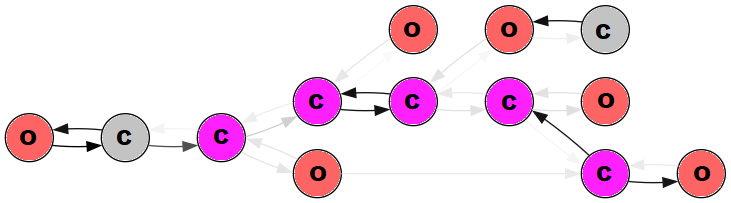}
        \caption{}
        \label{fig:5c}
    \end{subfigure}
    \begin{subfigure}{1\textwidth}
        \centering
        \includegraphics[width=0.8\linewidth]{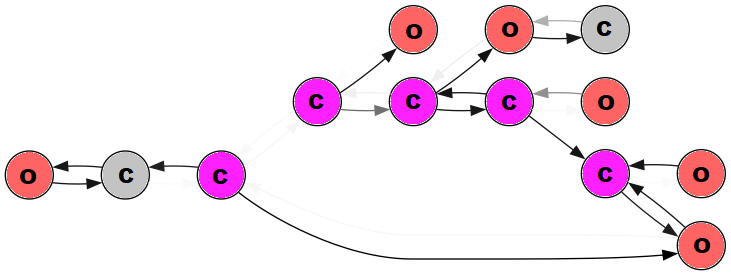}
        \caption{}
        \label{fig:5d}
    \end{subfigure}

    \caption{Comparative Representation of the outputs of GNNExplainer on the target  Molecule Name: \textit{6-(hydroxymethyl)-4-methoxyoxane-2,3,5-triol}, (a) Molecular Structure, (b) Node Feature Importance, (c) Output generated by the ECC model, (d) Output generated by the ECRGNN model. In this case, the target property considered is Aqueous Solubility. (The grayscale value of the arrows signifies the strength of the relationship between the two nodes, which directly correlates to the predicted property) }
    \label{fig:main}
\end{figure*}


\begin{figure*}
    \centering
    \captionsetup{justification=centerlast} 
    \begin{subfigure}{0.4\textwidth}
        \includegraphics[width=0.9\linewidth,  height=0.9\linewidth]{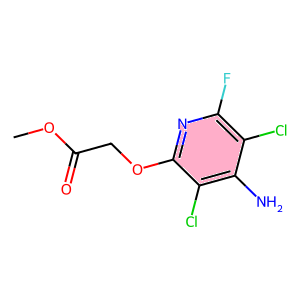}
        \caption{}
        \label{fig:6a}
    \end{subfigure}
    \begin{subfigure}{0.45\textwidth}
        \includegraphics[width=1\linewidth, height=0.7\linewidth]{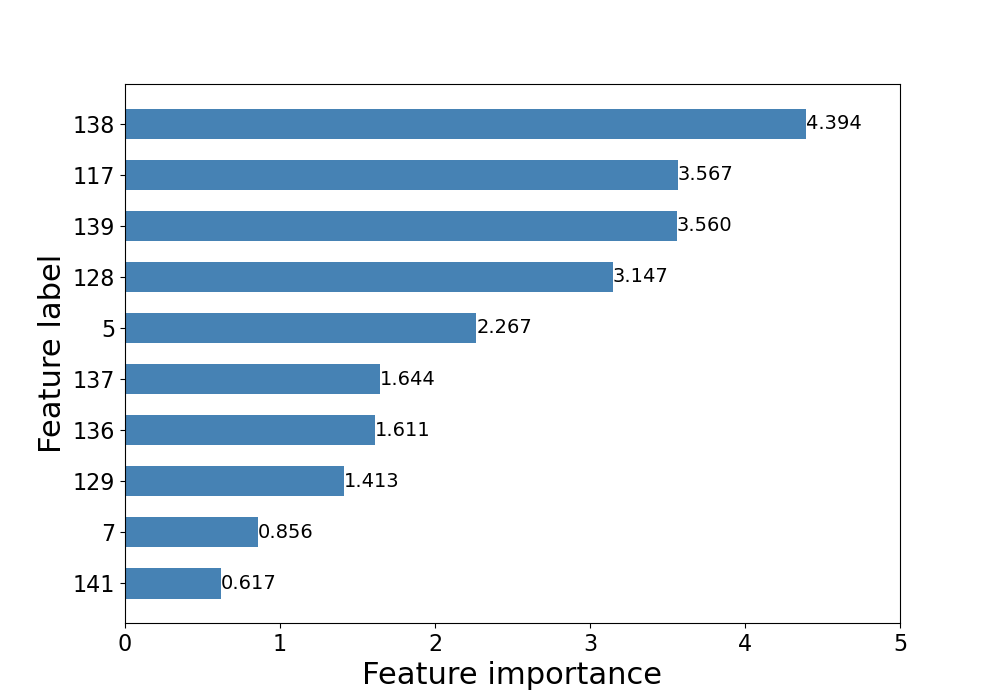}
        \caption{}
        \label{fig:6b}
    \end{subfigure}


    \begin{subfigure}{1\textwidth}
         \centering
        \includegraphics[width=1\linewidth]{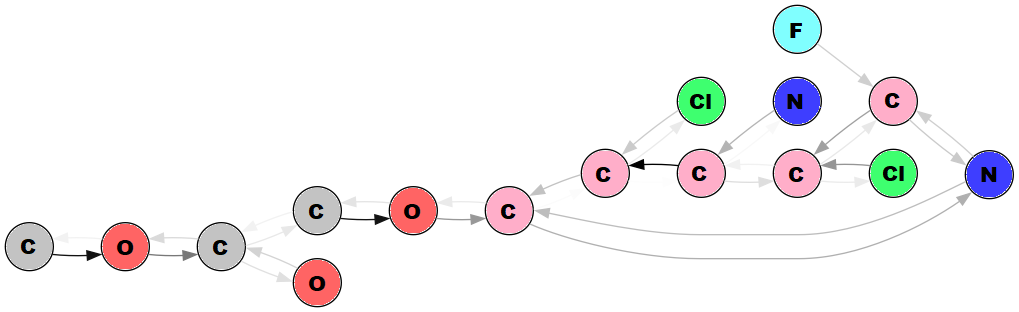}
        \caption{}
        \label{fig:6c}
    \end{subfigure}
    \begin{subfigure}{1\textwidth}
         \centering
        \includegraphics[width=1\linewidth]{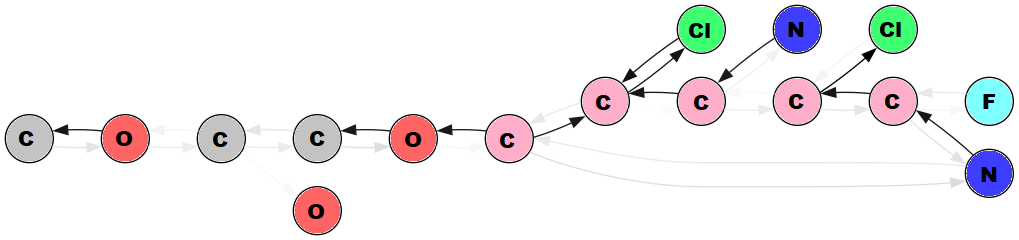}
        \caption{}
        \label{fig:6d}
    \end{subfigure}

    \caption{Comparative Representation of the outputs of GNNExplainer on the target Molecule Name: \textit{4-methoxy-1H-indazole} (a) Molecular Structure, (b) Node Feature Importance, (c) Output generated by the ECC model, (d) Output generated by the ECRGNN model. In this case, the target property considered is Aqueous Solubility. (The grayscale value of the arrows signifies the strength of the relationship between the two nodes, which directly correlates to the predicted property)}
    \label{fig:main}
\end{figure*}

The Figures \ref{fig:7c}, \ref{fig:8c}, \ref{fig:3c}, \ref{fig:4c}, \ref{fig:5c}, \ref{fig:6c} and Figures \ref{fig:7d}, \ref{fig:8d}, \ref{fig:3d}, \ref{fig:4d}, \ref{fig:5d}, \ref{fig:6d}  show the results of output of the GNNExplainer model on ECC model and ECRGNN models respectively.\cite{Cit_6} The actual structure of the molecules, along with the node importance, has been presented in Figure \ref{fig:7a}, \ref{fig:8a}, \ref{fig:3a}, \ref{fig:4a}, \ref{fig:5a}, \ref{fig:6a} and \ref{fig:7b}, \ref{fig:8b}, \ref{fig:3b}, \ref{fig:4b}, \ref{fig:5b}, \ref{fig:6b} respectively, which are the top left and top right correspondingly. The strength of the node and the edge connections are presented by varying the degrees of the black. The atoms, as well as the rings, are colour-coded to match the molecules.

Lipophilicity and aqueous solubility are properties that describe the degree of solubility of the input substance in non-polar and polar solvents, respectively. As the molecules are predominantly organic in nature, the feature importance indicates carbon and nitrogen to be the most important contributing atoms to the determination of these properties. Features such as formal charge and aromaticity are also deemed important features, as these affect the molecule's polarity by determining the molecule's final charge and polarity due to the probabilistic flow of electrons. This, in turn, directly relates to the fact that the more polar a molecule is, the higher the expected solubility in a polar solvent, while high polarity indicates a low solubility in non-polar solvents. An atom (node) being part of a ring is also identified as a feature of importance since that indicates the presence of a cyclic structure which can significantly impact a molecule's overall shape, electronic distribution, and thus its solubility properties. Cyclic compounds, especially those that are aromatic, tend to have de--localized $\pi$ electrons, which contribute to the molecule's stability and reactivity. This de-localization can affect the molecule’s polarity. Overall, the interplay of these factors—atom type, formal charge, aromaticity, and structural features such as being part of a ring—determines organic molecules' lipophilicity and aqueous solubility.

From the explanation graphs of lipophilicity and aqueous solubility, as shown in Figure \ref{fig:3d} and \ref{fig:6d}, one observation to be noted is that the ECRGNN model focuses intensively on connections that signify a high amount of electron flow. This is visible by the lighter (weaker) connections between bonds of similar atoms, like C-C bonds, and a darker (stronger) connection between bonds of Carbon with other atoms like Oxygen, Nitrogen, Fluorine, etc., which use their inherent electronegativity to draw electrons towards themselves, creating polar bonds. This polarization within the molecule is a key factor in determining its solubility characteristics. Bonds between atoms of different electronegativities (such as C-O, C-N, or C-F bonds) are more polar, making the molecules containing these bonds more likely to interact with polar solvents and less likely to interact with non-polar molecules using dipole-dipole interactions.

In the example of the molecule in Figure \ref{fig:7a} the performance of the ECRGNN model against the ECC \cite{Cit_6} model, it can be seen that the ECRGNN model accurately captures relations relevant to solubility that are missed by the ECC models, such as connections within the $CF_3$ subgroup, as well as connections within the neighbouring benzene ring. Additionally, the ECRGNN model captures other important polar bonds that affect lipophilicity, which the ECC model fails to capture. Thus, this ablation study provides concrete proof of the validity of the novel ideation. This also explains how having a balance of so-called electron-giving and electron-receiving groups results in a very high lipophilicity.

Figure \ref{fig:4c} and \ref{fig:4d} for the molecule (4-methylphenyl) urea depicts that the ECRGNN model captures heterogeneous bonds that are relevant in spreading the electron charge and making the molecule polar, while the ECC \cite{Cit_6} model highlights no such information. Thus, the ECRGNN model accurately captures the reasoning behind the high value of the aqueous solubility of the molecule.

\subsection{Variational Autoencoder (VAE)}
\label{VAE_block}
Numerous domains, including network system monitoring, proteomic studies, and drug molecule analysis, necessitate the classification of multiple labels based on graph-connected features such as molecule structures and network graphs. Datasets encoded in (SMILES) \cite{Cit_14} format, such as Tox21 \cite{Cit_34} and Sider \cite{Cit_34}, encapsulate valuable information pertaining to drug toxicities and other molecule parameters. An initial attempt to establish an enhanced benchmark involved a straightforward model consisting of a single layer of sigmoid neurons appended to the end of a novel GNN, its competitiveness was noteworthy. However, for multilabel prediction tasks, the efficacy of VAEs surpasses conventional models, inspiring a closer exploration of their workings.

VAEs encode input data into a probabilistic latent space and decode it to reconstruct the input. This reconstruction allows us to align the feature and label spaces, allowing for much more accurate predictions in the multi-label classification scenario.

The utilization of a VAE for capturing the probabilistic latent space presents significant advantages. Firstly, it grants enhanced control over the latent space. \cite{Cit_42} Notably, within various AE models, the label-encoder-decoder pathway often outperforms the feature-encoder-decoder pathway. Incorporating the VAE structure within the latent space aids in balancing the complexity of learning and aligning these distinct subspaces. Secondly, achieving smoothness within the latent space is a common objective. \cite{Cit_43} VAEs, being probabilistic models, inherently induce local-scale smoothness as the decoder interprets samples rather than specific embeddings. Thirdly, VAE models and their variations exhibit an ability to learn representations characterized by disentangled factors. \cite{Cit_44} When both latent spaces, pertaining to features and labels, learn these disentangled factors, it facilitates the alignment of these spaces and contributes significantly to the decoding process, augmenting the results.

\begin{figure}[hbt!]
    \centering
    \captionsetup{justification=centerlast} 
    \includegraphics[width=0.42\textwidth]{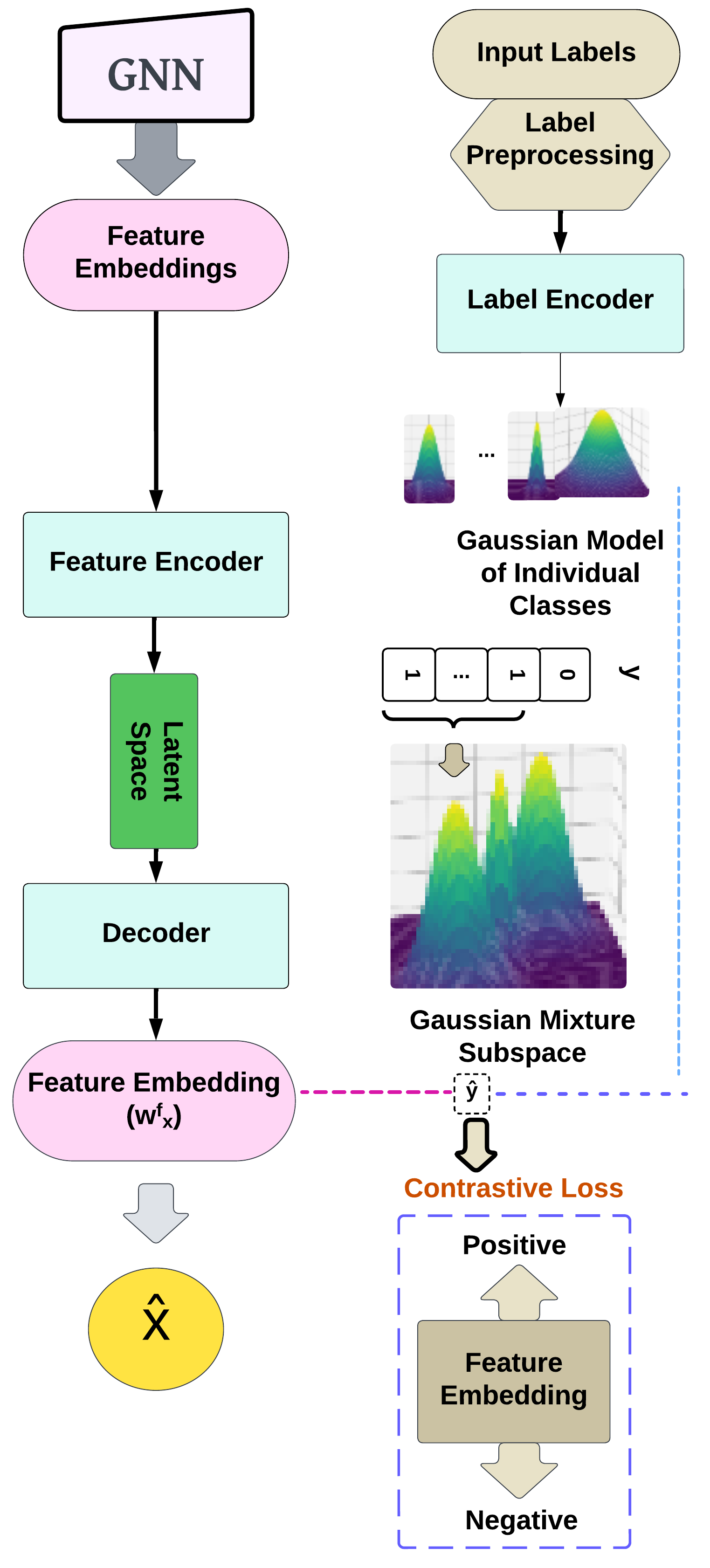}
    \caption{Architecture of VAE model used.}
    \label{fig:VAE_fig}
\end{figure}

While VAEs are often used for generative tasks, the Contrastive Gaussian Mixture Variational Autoencoder (C-GMVAE) \cite{Cit_60} is specially designed for multi-label classification (MLC). This model addresses the common challenges faced by vanilla VAEs \cite{Cit_66} such as over-regularization, posterior collapse, and label correlation in MLC tasks. The C-GMVAE leverages contrastive learning to strengthen label embedding learning and integrates a Gaussian mixture variational autoencoder (GMVAE) to synergistically learn a multimodal latent space.

The C-GMVAE model comprises two primary components: the Gaussian mixture VAE and the contrastive learning module. The Gaussian mixture VAE adheres to the conventional VAE framework, combining the posterior distribution with a parameter-free isotropic Gaussian prior. During training, it optimizes two essential losses: the KL divergence from the prior to the posterior and the distance between the reconstructed targets and the real targets. This component empowers the model to acquire a multimodal latent space, facilitating probabilistic modelling. A representative diagram of the C-GMVAE model has been presented in Figure \ref{fig:VAE_fig} to highlight its workflow in the multi-class multi-label classification setting.

A pivotal innovation within the C-GMVAE model is the contrastive learning module. This component employs a contrastive loss to enhance label embedding learning, effectively capturing label information in a supervised multi-label classification (MLC) scenario. By defining anchor, positive, and negative samples, the contrastive loss harnesses label information with great efficacy, eliminating the need for resource-intensive label correlation modules such as covariance matrices and GNNs. This innovative approach underscores the adaptability of contrastive learning, conventionally employed in self-supervised learning, to effectively capture label information in supervised MLC tasks.

\subsection{End to End Learning}
Since there are two problems to be solved, regression and classification, both will have their own independent learning schemes. The entire GNN backbone can be represented as $ F_n: G(n) \rightarrow m_f $, where $F_n$ is the GNN function that generates the molecular fingerprint/embeddings $(m_f)$. The MLP as defined in Sec.\ref{exap_gan_algorithm} is automatically integrated in the GNN function which allows for backpropagation. The embeddings are then used for the general downstream tasks. For the regression task, the task is defined in the format $MLP(m_f) \rightarrow \text{output prediction}$.

\begin{figure*}[hbt!]
    \centering
    \captionsetup{justification=centerlast} 
    \includegraphics[width=0.8\textwidth]{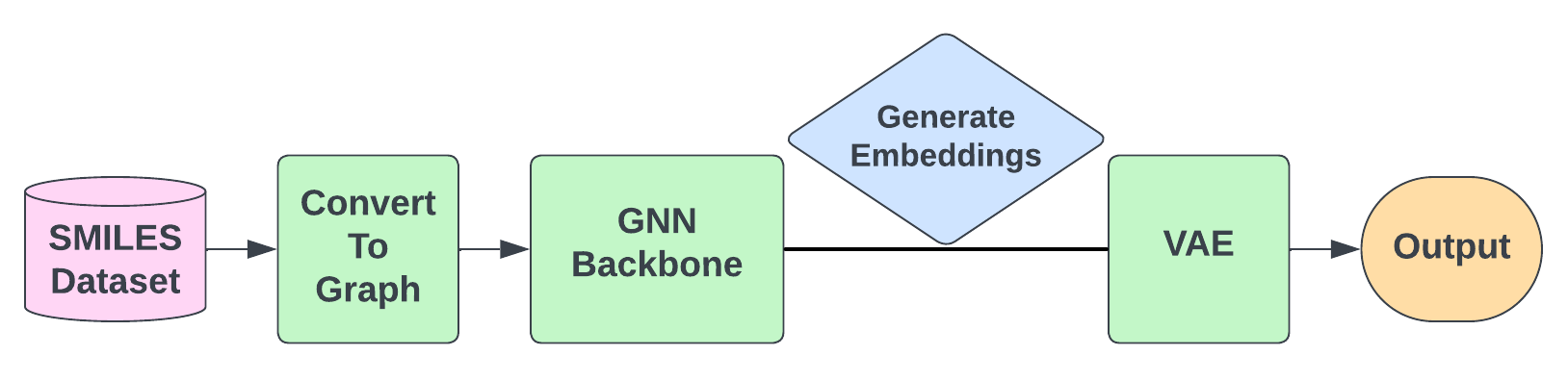}
    \caption{Diagrammatic representation of end-to-end learning.}\label{fig:end_to_end_learning_fig}
\end{figure*}

The multilayer perceptron is applied to molecular embeddings generated by the GNN. Since the entire updation takes place in a single setting and is bound by a single computation graph, proper end-to-end learning takes place. For the multi-class multi-label classification setting, a diagrammatic representation of the workflow has been represented in Figure \ref{fig:end_to_end_learning_fig}. As can be seen, the SMILES data is first converted to a graph database by PyTorch-geometric \cite{Cit_22}, with the features of the nodes and edges being extracted by Rdkit \cite{Cit_64}. The dataset is then passed through the GNN as represented above to generate $m_f$.The VAE part of the execution can be represented as $VAE(m_f) \rightarrow \hat{p}$ and $\sigma(\hat{p}) \rightarrow \text{predicted vector}$.

In the current scenario, an MLP is used as the encoder and the decoder of the VAE; however, it can be easily replaced with much more sophisticated VAE methods such as \cite{Cit_65}. The VAE generates a probability vector, which will give us the predicted vector when passed through the sigmoid function. In both the regression and classification scenarios, all the entities are attached to an individual computation graph. Thus, this allows for an end-to-end manner of training where the generated molecular fingerprint is dependent on all the entities, and backpropagation is allowed throughout the computation graph.

\section{Model Setup and Training}
For the model setup, the ECRGNN backbone described in Sec. \ref{GNN_block} is used for both the regression and classification tasks. All the hyper-parameters of the graph convolutions and the fully connected layers remain the same. In the case of regression property prediction, Huber loss, a combination of MSE and MAE, was used to account for the presence of outliers. The resultant loss generated is converted into an exponential loss function.

The cross-entropy loss function has been used in the case of Multi-class/Multi-label Classification datasets. The ECRGNN backbone, along with a sigmoid layer of one-hot encoded classes, has been used. To prevent the class imbalance from affecting the results, a self-adaptive weighted loss function controlled by Adam optimizer \cite{Cit_59} is implemented. This prevents some classes from becoming over-dominant, thus increasing the F1 score.

The ECRGNN backbone, in conjunction with VAE, has been used to establish an end-to-end pipeline for training. The hyperparameters of the GNN backbone remain the same, and those of the VAE have been described in Sec. \ref{VAE_block}. To explain the ECRGNN-VAE model's efficacy, we also present comparative results of the ECRGNN backbone with the last layer as a sigmoid activation function.

Adam optimizer and Cosine annealing scheduler with warm restarts have been used in both regression and classification cases. The batch size is 64, which is found to be the optimum, and we have run the model for a maximum of 400 epochs. The learning rate is initialized with a value of 0.001 and a weight decay of 5e-4. An early stopping criterion has a patience value of 20 epochs. The entire dataset is divided into 85\% Train set, 10 \% Test set, and 5 \% Validation set, and this has been maintained across all the datasets.

\section{Regression}
This section demonstrates the experiments on the regression. The main metrics used for quantifying the results have been presented in Table \ref{tab:Table_5}. A detailed description of the data preprocessing steps for all formats of the dataset is given below. For a visual representation of the results, parity plots are presented 
\subsection{Dataset Description}
There are several properties, such as boiling point, melting point, solubility, etc, which are not readily available for use in the case of organic as well as inorganic compounds. These properties take time and resources to be found, and the extraction of group contributions to these molecules is not easy. For novel compounds, these values are not readily available. Thus, the efficacy of the GNN model to directly extract the property from its SMILES \cite{Cit_16} representation is accurately demonstrated based on some of the example properties.

For regression problems, datasets are referred from Asheri et al. \cite{Cit_33}, which contain over 24,259 molecular compounds and their distinct properties. In order for the model. Among them properties such as Boiling Point \cite{Cit_33}, Acentric factor \cite{Cit_33}, LogWs \cite{Cit_33}, LogP \cite{Cit_33}, Melting Point \cite{Cit_33}, LD50\cite{Cit_33} have been chosen. Apart from that, results on standard benchmark datasets such as the LipO (lipophilicity) dataset \cite{Cit_34} have been show-cased.
\subsection{Regression Results}
The ECRGNN method demonstrates substantial advancements over prior benchmarks in the field. Building upon the foundational work of Asheri et al. \cite{Cit_33} and Rittig et al. \cite{Cit_6}, where R\textsuperscript{2} score of 0.85 has been reported, the revised GNN model has achieved a notable enhancement in performance in the Boiling point dataset\cite{Cit_33} both in training and testing phases. As a benchmark, in the test phase, the ECRGNN showcased a significant improvement, achieving a 10.1\% increase in R\textsuperscript{2} score on average over the single GNN model \cite{Cit_6} and 3.41\% increase over their ensembled GNN model \cite{Cit_6}, thus surpassing the previous state-of-the-art results.

\begin{table*}[hbt!]
\centering
\captionsetup{justification=centerlast} 
\caption{Prediction Results from Graph Regression. }
\label{tab:Table_5}
\small
\resizebox{\textwidth}{!}{%
\begin{tabular}{ccccccc}
\hline
DATASET & DATAPOINTS & R\textsuperscript{2} & MAE & MSE & MAPE & RMSE \\
\hline
Boiling point & 507 & 0.934 & 12.375 & 375.83 & 2.72\% & 19.38 \\
Melting Point & 686 & 0.816 & 30.915 & 1895.08 & 9.09\% & 43.53 \\
Acentric factor & 172 & 0.843 & 0.0089 & 0.0002 & 1.48\% & 0.0142 \\
LogWs & 247 & 0.883 & 0.0891 & 0.0167 & 2.25\% & 0.1294 \\
Lipo & 420 & 0.781 & 0.0745 & 0.0100 & 2.6\% & 0.1004 \\
LogP & 1202 & 0.910 & 0.0256 & 0.0011 & 6.99\% & 0.0334 \\
Acid Dissolution & 160 & 0.766 & 0.1245 & 0.0355 & 1.64\% & 0.1885 \\
LD50 & 472 & 0.606 & 0.0591 & 0.0056 & 2.20\% & 0.0749 \\
\hline

\end{tabular}%
}
\end{table*}

To verify the ECRGNN's performance, results of various other datasets of Asheri et al. \cite{Cit_33} are cross-referenced, and the ECRGNN obtained competitive or better metrics on these sets. It is observed that the model performs extremely well on those sets where the number of data points is comparatively larger than others. This is inherently due to the fact that GNN models require a comparatively vast amount of data to properly generalize.

Comparing the results of the ECRGNN on benchmark datasets, such as Lipophilicity \cite{Cit_34}, an RMSE score of 0.1 is obtained on average on the test set, which is a massive improvement from \cite{Cit_8} and \cite{Cit_7} which are currently the top benchmarks in the Lipophilicity dataset \cite{Cit_34} and reports a RMSE score of 0.587 along with pretraining on additional dataset. It is to be noted that no externally added supplementary data has been used for modelling, which could have helped in generalization. The results of this paper not only live up to the standards of the current benchmark models but also show the robust generalization capability of the model across all sorts of datasets.

\begin{figure*}[hbt!]
     \captionsetup{justification=centerlast} 
     \centering
     \begin{subfigure}[b]{0.3\textwidth}
         \centering
         \includegraphics[width=\textwidth,height=0.8\textwidth,keepaspectratio]{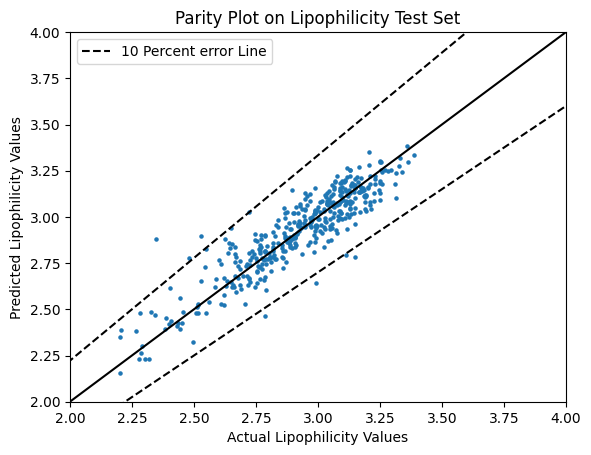}
         \caption{}
         \label{fig:Lipo}
     \end{subfigure}
     \hfill
     \begin{subfigure}[b]{0.3\textwidth}
         \centering
         \includegraphics[width=\textwidth,height=0.8\textwidth,keepaspectratio]{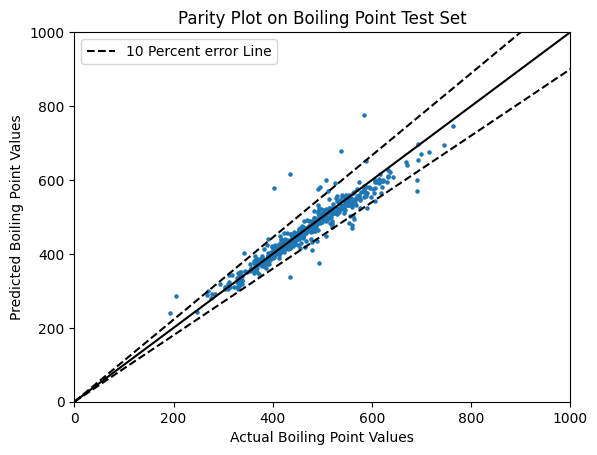}
         \caption{}
         \label{fig:Boiling}
     \end{subfigure}
     \hfill
     \begin{subfigure}[b]{0.3\textwidth}
         \centering
         \includegraphics[width=\textwidth,height=0.8\textwidth,keepaspectratio]{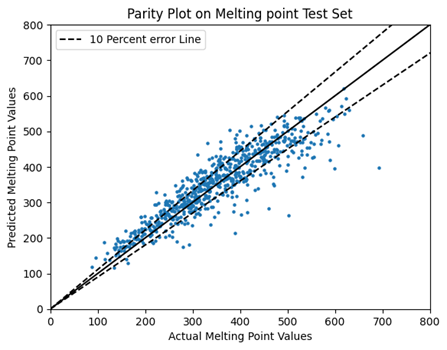}
         \caption{}
         \label{fig:Melting_p}
     \end{subfigure}
     
     \vspace{1em}
     
     \begin{subfigure}[b]{0.3\textwidth}
         \centering
         \includegraphics[width=\textwidth,height=0.8\textwidth,keepaspectratio]{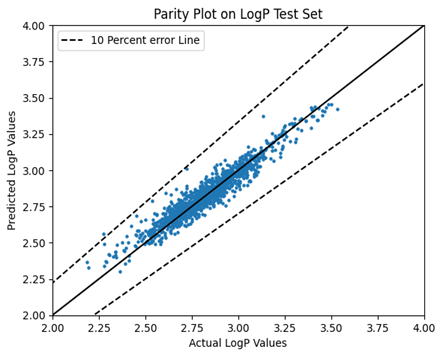}
         \caption{}
         \label{fig:LogP}
     \end{subfigure}
     \hfill
     \begin{subfigure}[b]{0.3\textwidth}
         \centering
         \includegraphics[width=\textwidth,height=0.8\textwidth,keepaspectratio]{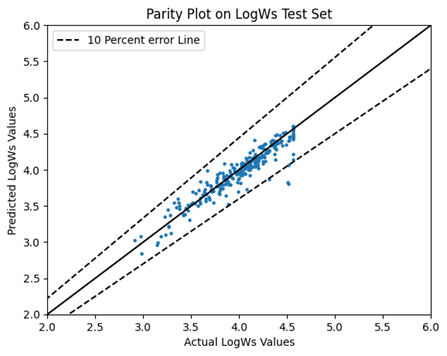}
         \caption{}
         \label{fig:Logws}
     \end{subfigure}
     \hfill
     \begin{subfigure}[b]{0.3\textwidth}
         \centering
         \includegraphics[width=\textwidth,height=0.8\textwidth,keepaspectratio]{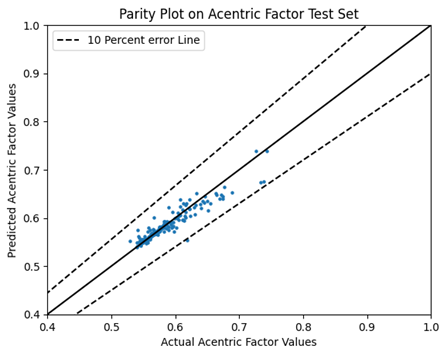}
         \caption{}
         \label{fig:acentric}
     \end{subfigure}
     
     \vspace{1em}
     
     \begin{subfigure}[b]{0.3\textwidth}
         \centering
         \includegraphics[width=\textwidth,height=0.8\textwidth,keepaspectratio]{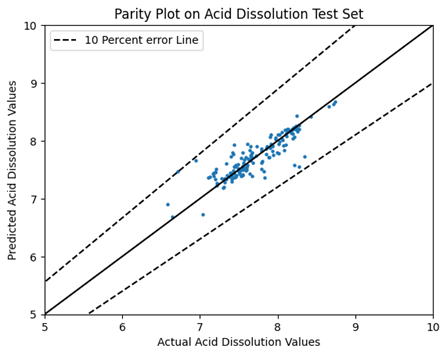}
         \caption{}
         \label{fig:acid}
     \end{subfigure}
     \hspace{0.1\textwidth}
     \begin{subfigure}[b]{0.3\textwidth}
         \centering
         \includegraphics[width=\textwidth,height=0.8\textwidth,keepaspectratio]{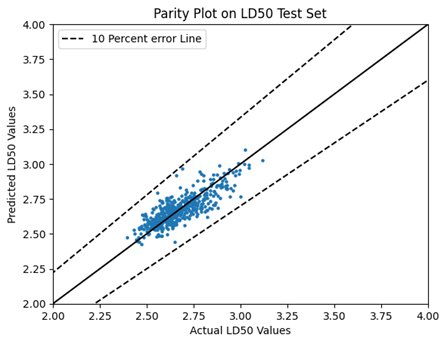}
         \caption{}
         \label{fig:ld50}
     \end{subfigure}
     \caption{Parity plot of various datasets of \cite{Cit_33} on their test sets as presented - (a) Lipophilicity (b) Boiling Point (c) Melting Point (d) LogP (e) LogW (f) Accentric Factor (g) Acid Dissolution factor (h) LD50 dataset.}
     \label{fig:6}
\end{figure*}

To further highlight the distribution of points and visualize the results, Parity plots of the test set have been provided for all the datasets in Figure \ref{fig:6} to show the grouping of the results on the test set. The actual experimental values have been plotted in the x-axis whereas the predicted values have been plotted in the y-axis. From the plots, it can be seen that the data points lie mostly within the ±10\% error line, even for large datasets like Boiling Point, LogP, and LD50, as shown in Figure \ref{fig:Boiling}, \ref{fig:Logws}, \ref{fig:ld50}. Since none of the data points have been excluded, the parity plots show that even for cases that are outliers, the model performs well, and the data points lie within the acceptable limits. However, it can be seen that the data points for melting points are not that closely packed. This can be reflected in the R\textsuperscript{2} score as well as the model's failure to generalize much of the dataset. Apart from the Melting Point dataset \ref{fig:Melting_p}, more than 98\% of the dataset for other datasets fall below the 10\% error range.

Additionally, the method's success on the LogW, LogP, and Acentric factor datasets highlights its adaptability and generalizability on different datasets. For a clear comparison across all datasets as well as with current benchmark models, all forms of results such as MAE, MSE, and R\textsuperscript{2} scores have been recorded for each dataset.

\section{Classification}
Graph classification plays a pivotal role in the prediction of behaviours of different molecules and has various applications in the domains of drug discovery \cite{Cit_35}, the generation of new materials \cite{Cit_36}, and identifying the intrinsic properties of different compounds \cite{Cit_27}. Given the superior performance of the new GNN model in regression scenarios, subjecting it to classification tasks is imperative, thereby affirming its robustness in discerning intricate patterns across data of different tasks. Multi-label multi-class classifications are performed. It is a domain where past experiments have been sparse, proving the model’s ability to handle diverse and complex tasks.

In this study, a comparative analysis has been done between a conventional methodology involving GNN feature embeddings fed into a multi-neuron sigmoid layer and a novel approach with the GNNs as a backbone to a VAE model, as expounded in Sec. \ref{VAE_block}. The latter method facilitates end-to-end learning of the GNN and VAE blocks simultaneously. A comprehensive comparison is performed to highlight the results and advantages of the novel approach over the conventional approach.
\subsection{Dataset Description}
For the classification setting, the focus has been on datasets with multiple classes or sub-tasks, and each data point is associated with one of those labels. With this setting in mind, popular datasets have been chosen to benchmark the ECRGNN's performance against other state-of-the-art algorithms. A detailed description of the datasets used has been presented. The Tox21\cite{Cit_34} dataset includes a set of chemical compounds examined for toxicity across 12 different biological targets. With over 8,000 compounds, it provides valuable information for predicting the toxicity and potential adverse effects of various chemicals. Another dataset such as SIDER \cite{Cit_34} provides a helpful resource about drug side effects. It gathers info from different sources like clinical trials and medical reports. It helps check if drugs are safe and understand side effects. There are 27 tasks in the dataset that provide useful insights into the possible side effects of different drugs. Furthermore, the classification models have been tried on higher-dimensional label space methods like Toxcast \cite{Cit_34}, which consists of more than 8000 compounds and 617 labels. Similarly, with PCBA dataset \cite{Cit_34}, generated by high throughput screening for biologically active molecules. Like Toxcast, this has a 128-dimensional label space and over 400,000 compounds.
\subsection{Dataset Processing}
\label{Class_dat}
Regrettably, several SMILE-encoded datasets earmarked for multilabel multiclass classification exhibited anomalies such as missing labels and vacant entries. While rows with empty entries could be easily excluded, coping with partially missing labels posed a more intricate challenge. Imputing these labels was deemed suboptimal, as it risked compromising the authentic intricacies of the real-world data. Consequently, the preprocessing unfolded in two distinct phases. Firstly, entries with over 50\% missing labels were expunged from the dataset. Despite a reduction in data volume, this step ensured the preservation of a certain qualitative threshold, retaining crucial representative elements. Secondly, missing data was imputed with a constant value (initially 0) to address the unavoidable necessity. This step, however, introduced some data corruption and exacerbated the existing class imbalance. Notably, the SIDER \cite{Cit_34} dataset remained pristine, devoid of such defects, and was used directly. In contrast, Tox21\cite{Cit_34} and ToxCast \cite{Cit_34} datasets suffered more significantly from this issue.

\begin{figure*}[hbt!]
\captionsetup{justification=centerlast} 
\centering
     \begin{subfigure}[b]{0.48\textwidth}
         \centering
         \includegraphics[width=\textwidth]{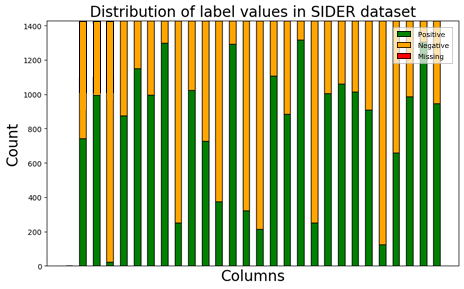}
         \caption{}
         \label{fig:sid}
     \end{subfigure}
     \hfill
     \begin{subfigure}[b]{0.48\textwidth}
         \centering
         \includegraphics[width=\textwidth]{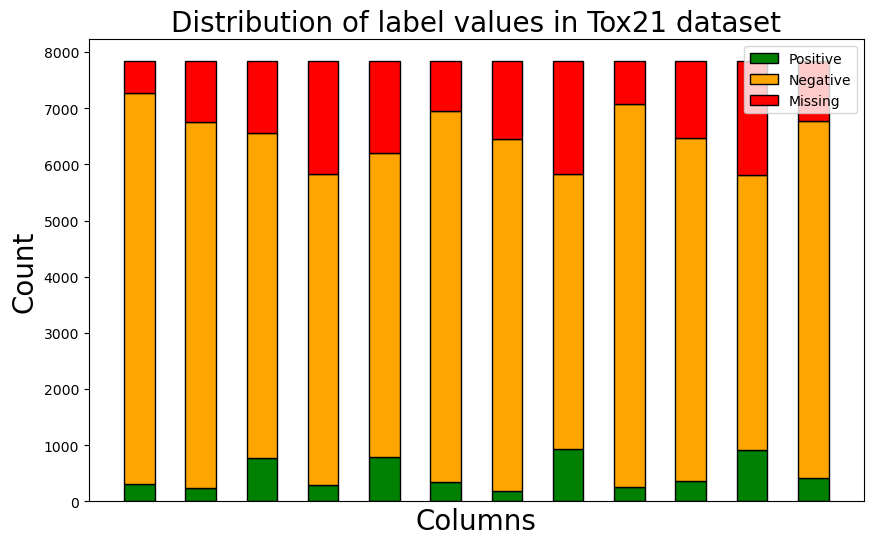}
         \caption{}
         \label{fig:tox21}
     \end{subfigure}
     \hfill
     \begin{subfigure}[b]{0.48\textwidth}
         \centering
         \includegraphics[width=\textwidth]{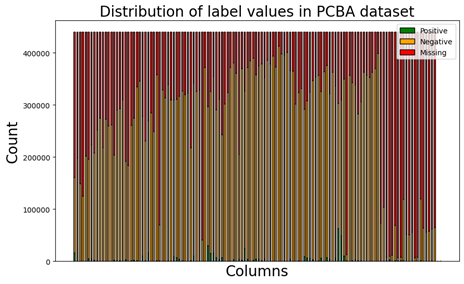}
         \caption{}
         \label{fig:PCBA}
     \end{subfigure}
     \hfill
     \begin{subfigure}[b]{0.48\textwidth}
         \centering
         \includegraphics[width=\textwidth]{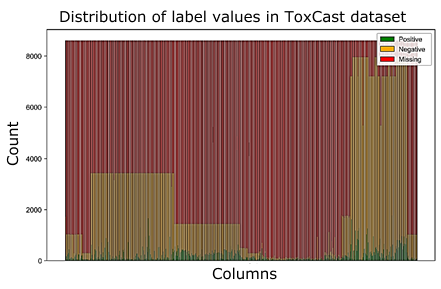}
         \caption{}
         \label{fig:toxcast}
     \end{subfigure}
     \hfill

\caption{Statistical plot of Class imbalance as well as missing values of the datasets (a) SIDER, (b) Tox21, (c) PCBA,  (d) ToxCast.}
\label{fig:stat}
    \end{figure*}%

Figure \ref{fig:stat} is a statistical analysis of the imbalances in the dataset with respect to the labels. As can be seen, there is a huge mismatch between the 1’s and 0’s of all the datasets. One of the major problems in these datasets is the presence of a huge number of NAN/missing values. This problem, although not present with sider, is predominant in the other datasets such as Tox21\cite{Cit_34}, PCBA \cite{Cit_34}, and especially ToxCast \cite{Cit_34}. Since imputing all the values with 0 was not deemed ideal, experiments were conducted by giving NAN values a separate class to give them proper representation, the results of which have been discussed. Thus, the problem formulation has been presented for both binary classes as well as multi-class multi-label classification.

\subsection{Results of ECRGNN+sigmoid}
Before the ECRGNN + VAE model, ablation studies were performed using the base ECRGNN model. Table \ref{tab:Table_6} describes the results obtained directly from the ECRGNN backbone, with the last layer being the sigmoid layer. As can be seen from the table, the results are not satisfactory enough with just a single sigmoid layer. The scores obtained, however, surpass those obtained from the ECRGCN \cite{Cit_46} backbone, which further highlights the model’s effectiveness. 

\begin{table}[hbt!]
\small
\centering
\captionsetup{justification=centerlast} 
\caption{Results of GNN+sigmoid method.}
\label{tab:Table_6}
\begin{tabular}{ccccc}
\hline
DATASET & Macro-AUCROC & Ha & Total Accuracy & Micro-F1 \\
\hline
SIDER & 0.6234 & 0.656 & 0.056 & 0.6956 \\
Tox21 & 0.5871 & 0.811 & 0.46 & 0.6133 \\
\hline

\end{tabular}%

\end{table}

As can be seen from the results, the primitive method of directly producing the labels is not able to be accurately represented by the ECRGNN model, and the accuracy is very low. Thus, for multi-class multi-label settings, different options for accurately representing the labels to the embeddings, the ECRGNN+VAE model, has been introduced. A major advantage of this is that the VAE can be directly attached to the ECRGNN backbone.

\subsection{Results of ECRGNN+VAE model}
The predominant metric for multilabel multiclass classification is the micro-AUC-ROC score, as it is robust against class imbalance in different labels. Previous studies, specifically Elembert-v1\cite{Cit_26}  and BioAct-Het\cite{Cit_27}, utilized a single-model-per-label approach in their work on the Tox21\cite{Cit_34} and Sider \cite{Cit_34} datasets. While effective for molecular property prediction and drug discovery, this approach proves impractical in high-dimensional label-space scenarios, demanding substantial time and resources. In contrast, the GNN+VAE model for multilabel multiclass classification predicts all labels simultaneously, significantly reducing the resources required for both training and inference. Notably, outperforming the state-of-the-art results reported in Table \ref{tab:Class_VAE} on the sider dataset, the model achieves an AUC-ROC score of 93.26\%, surpassing their reported 91\% average. There is also a discrepancy to be noted. The BioAct-Het \cite{Cit_27}  model reports metrics on a subset of labels, while ECRGNN provides predictions for all labels in the dataset. On the tox21 dataset, ECRGNN achieves an AUC-ROC score of 84\%, somewhat lower than Elembert's\cite{Cit_26} mean score of 96\%, attributed to the smaller number of labels in the Tox21 dataset contributing to better learning of single-model-per-label systems. Despite this, the approach proves to be a competitive and resource-efficient alternative.

\begin{table*}[hbt!]
\small
\centering
\captionsetup{justification=centerlast} 
\caption{Predictions for (Imputation 0) Classification.
}
\label{tab:Class_VAE}
\resizebox{\textwidth}{!}{%
\begin{tabular}{cccccc}
\hline
DATASET & Macro-AUCROC & Ha & Total Accuracy & Micro-F1 & Mico-AUCROC \\
\hline
SIDER & 0.8321 & 0.7812 & 0.0933 & 0.8560 & 0.9326 \\
Tox21 & 0.7476 & 0.92 & 0.31 & 0.7522 & 0.8478 \\
ToxCast & 0.587 & 0.67 & 0.04 & 0.828 & 0.7821 \\
PCBA & 0.6741 & 0.735 & 0.035 & 0.787 & 0.7521 \\
\hline

\end{tabular}%
}

\end{table*}

On distinct 85-5-10 train-validation-test splits, the ECRGNN demonstrates robust performance on sider, achieving a 76.7\% hamming accuracy. Notably, the model excels on Tox21 with an impressive 60\% accuracy and a consistent 92\% hamming accuracy. The observed discrepancy is due to the lower label dimension in Tox21, contributing to the model's optimal performance in this setting.

On the PCBA dataset, the ECRGNN underperforms with an AUC score of 75\%, quite lower than the benchmark set \cite{Cit_29} at 87\%. The model showcases competitive performance in the expansive ToxCast dataset featuring an impressive 617 task labels, achieving a notable 78.2\% 

The versatility and adaptability of ECRGNN are evident in its consistent performance across these diverse datasets. The substantial enhancements in Micro-AUC-ROC scores, particularly in Sider, and the model’s ability to provide competitive results even for a high dimensional label space like that of ToxCast, underscore not only the efficacy of the approach but also its potential for accurate and comprehensive multilabel multiclass classification tasks in various domains. This, in turn, affirms the robustness and generalizability of the new GNN+VAE model. While ECRGNN excels at some tasks, some other datasets have better methods to deal with them individually. The upside to this is that ECRGNN performs consistently and competitively on all SMILE-encoded classification datasets with SOTA or near-SOTA performance.

MicroAUC is favoured over macroAUC in multilabel classification because it provides a balanced evaluation of model performance across all labels. Unlike MacroAUC, MicroAUC considers the contributions of each label equally, making it robust to class imbalance and sensitive to rare classes. By aggregating true positive and false positive rates, MicroAUC offers a comprehensive measure of the model's discriminative ability, making it well-suited for assessing overall performance in multilabel classification tasks.

Apart from the AUCROC score, other major criteria include accuracy. Though the AUCROC is an important metric, it highlights only the model’s discriminatory power and provides no information about the accuracy of the prediction. 

In a multi-label setting, where instances can belong to multiple classes simultaneously, hamming accuracy $(H_{a})$ as described in Eq. \ref{4_eqn} emerges as a more pertinent evaluation metric compared to traditional accuracy. Unlike accuracy, which collectively evaluates the correctness of predictions on all labels, Ha measures the fraction of correctly predicted labels per instance. $H_a$ accounts for the inherent complexity of multilabel classification tasks by assessing the model's ability to independently predict each label, offering a more comprehensive assessment of its predictive capabilities. $H_a$ is given by

\begin{equation}\label{4_eqn}
H_a = \frac{1}{n} \sum_{i=1}^{n} \frac{1}{m} \sum_{j=1}^{m} \delta(y_{ij} = \hat{y}_{ij}), \tag{4}
\end{equation}

where, $n$ is the number of samples, $m$ is the number of labels, $y_{ij}$ is the true value of the j\textsuperscript{th} label for the i\textsuperscript{th} sample, is the Predicted value of the j\textsuperscript{th} label for the i\textsuperscript{th} sample 

\begin{figure}[hbt!]
\centering
\captionsetup{justification=centerlast} 
\includegraphics[width=0.48\textwidth]{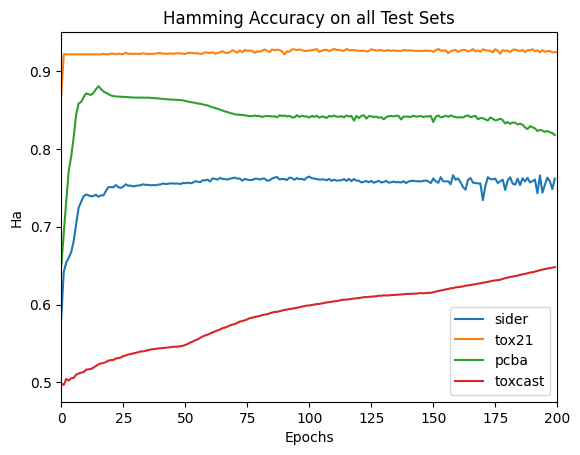}
\caption{Plot of Hamming Accuracy vs Epochs to observe the improvement in accuracy of the model.}
\label{fig:HA_fig}
\end{figure}

A representation of the Hamming Accuracy of the model on the test set has been plotted vs the epochs. This shows the model performance with epochs over the dataset. In Figure \ref{fig:HA_fig}, both total accuracy scores and Hamming accuracy scores have been presented. The reason for the importance of Ha being more than the total accuracy is due to their formulation as described in Eq. \ref{4_eqn}. The total accuracy is stricter in the sense that it penalizes the entire set of labels of a data point in case of a mismatch in multi-label settings. Meanwhile, the HA considers each individual label in a label set as a single entity to be compared.

Apart from the normal prediction, the idea of setting up separate labels for the missing values instead of directly imputing the values is proposed. This is highly effective for those datasets where the number of missing values is quite high in proportion to the other classes, such as PCBA and ToxCast, as can be seen from the statistical analysis in Figure \ref{fig:stat} in Sec.\ref{Class_dat}. The problem formulation now becomes a 3-class problem consisting of labels 0, 1, and 2, which are then one-hot encoded in the vector space. The reason for adding a separate label is that some of these missing labels may be inherently correlated with the property of the molecule, i.e., maybe those tests cannot be performed on that molecule to find the property.

\begin{table*}[hbt!]
\small
\centering
\captionsetup{justification=centerlast} 
\caption{Predictions for (Imputation 2) Classification.  }
\label{tab:Table_7}
\resizebox{\textwidth}{!}{%
\begin{tabular}{ccccc}
\hline
DATASET & Macro-AUCROC & Ha & Total Accuracy & Mico-AUCROC \\
\hline
Tox21 & 0.812 & 0.8806 & 0.64 & 0.8478 \\
ToxCast & 0.7491 & 0.8069 & 0.063 & 0.8011 \\
PCBA & 0.7577 & 0.735 & 0.203 & 0.7967 \\
\hline

\end{tabular}%
}
\end{table*}

From the results, as shown in Table \ref{tab:Table_7}, a major increase in R\textsuperscript{2} values can be seen for Tox21 as well as ToxCast datasets. Furthermore, a substantial increase in total accuracy as well as Hamming accuracy can be seen. Since nothing specific has been provided on the kind of preprocessing that has been done in the papers \cite{Cit_31}, from the results, it can be concluded that adding a separate class for the missing values seems more appropriate than imputing them.

\section{Discussion}
To evaluate the effectiveness of ECRGNN, the results are benchmarked with the current state-of-the-art models:  Elembert\cite{Cit_35, Cit_26}, GTOT\cite{Cit_31}, BioAct-Het\cite{Cit_27}, GMT\cite{Cit_28}. Benchmarking was done for both classification and regression datasets.

\subsection{Comparatative Study}

\begin{table*}[hbt!]
\centering
\captionsetup{justification=centerlast} 
\caption{Comparison for Regression.}
\label{tab:Table_9}
\small
\resizebox{\textwidth}{!}{%
\begin{tabular}{ccccccc}
\hline
DATASET & METRIC& ECRGNN& GMT+GCN& GMT + ECRGNN& Elembert& GTOT\\
\hline
Boiling point & RMSE& 19.38& 27.35 & \textbf{18.19} & 24.22& 20.78\\
Melting Point & RMSE& 43.53& 51.98 & 37.36 & 38.76& \textbf{34.21}\\
Lipo & RMSE& 0.1004& 0.5411 & \textbf{0.1003} & 0.1162& 0.1173 \\
LogP & RMSE& \textbf{0.0334}& 0.1360 & 0.0789 & 0.1450& 0.0926\\
\hline
Boiling point & MAE& 12.375& 17.14 & \textbf{12.11} & 13.89& 12.82 \\
Melting Point & MAE& 30.915& 42.031 & 26.09 & \textbf{25.65} & 26.17\\
Lipo & MAE& 0.0745& 0.1192 & \textbf{0.0621} & 0.0880& 0.079 \\
LogP & MAE& 0.0256& 0.0721 & \textbf{0.0249} & 0.0245& 0.0251\\
\hline
Boiling point & MAPE& 2.72\%& 3.81\% & \textbf{2.65\%} & 2.77\%& 2.74\%\\
Melting Point & MAPE& 9.09\%& 11.55\%& 8.67\% & \textbf{8.45\%} & 8.77\%\\
Lipo & MAPE& 2.6\%& 4.2\% & \textbf{2.52\%}  & 2.9\%& 2.71\%\\
LogP & MAPE& 6.99\%& 13.31\% & \textbf{6.91\%} & 8.25\%& 8.27\%\\
\hline

\end{tabular}%
}

\end{table*}

Table \ref{tab:Table_9} shows a comparative study of regression experiments by various models to predict Boiling Point \cite{Cit_33}, Melting Point \cite{Cit_33}, Lipophilicity \cite{Cit_33} and LogP \cite{Cit_33}. For accurate comparison, RMSE, MAE, and MAPE are presented. ECRGNN outperforms the other state-of-the-art models. If the pooling layer of the ECRGNN is replaced with GMT \cite{Cit_28}, the performance increases across many datasets. In the case of the Melting Point dataset \cite{Cit_33}, it has been noticed that Elembert \cite{Cit_35} and GTOT \cite{Cit_31} outperform the ECRGNN model. This may be because the Melting Point \cite{Cit_33} has a much larger training set than others, and since transformers require a huge amount of data to work with, it follows that these model's performance will improve with dataset size. However, it may be noted that ECRGNN can perform effectively even with small- and large-scale datasets.\\

\begin{table*}
\small
\centering
\captionsetup{justification=centerlast} 
\caption{Comparison for Classification (0 imputation)}
\label{tab:Table_10}
\resizebox{\textwidth}{!}{%
\begin{tabular}{ccccccc} 
\hline 
DATASET & Metric & ECRGNN & GMT+GCN & GMT+ECRGNN & Elembert & GTOT \\ 
\hline 
Sider & AUCROC & 0.8321 & 0.8123 & \textbf{0.8412} & 0.778 & 0.6382 \\ 
Tox21 & AUCROC & 0.7476 & 0.5389 & 0.66856 & \textbf{0.872} & 0.7236 \\ 
ToxCast & AUCROC & 0.587 & 0.6021 & 0.5568 & \textbf{0.739} & 0.6782 \\ 
\hline 
Sider & HA & \textbf{0.7812} & 0.7491 & 0.75421 & 0.623 & 0.5146 \\ 
Tox21 & HA & 0.92 & 0.891 & 0.926 & 0.918 & \textbf{0.9286} \\ 
ToxCast & HA & 0.67 & 0.9621 & 0.9706 & 0.961 & \textbf{0.9768} \\ 
\hline 
\end{tabular}%
}
\end{table*}

\begin{table}[hbt!]
\small
\centering
\captionsetup{justification=centerlast} 
\caption{Comparison for Classification (2 imputation)}
\label{tab:Table_11}
\resizebox{\textwidth}{!}{%
\begin{tabular}{ccccccc}
\hline
DATASET & Metric & ECRGNN & GMT+GCN & GMT+ECRGNN & Elembert & GTOT\\
\hline
Tox21 & AUCROC & \textbf{0.812} & 0.6023 & 0.6549 & 0.745 & 0.7613 \\
ToxCast & AUCROC & \textbf{0.7491} & 0.6314 & 0.6012 & 0.681 & 0.7013 \\
\hline
Tox21 & HA & \textbf{0.8806} & 0.7823 & 0.8706 & 0.798 & - \\
ToxCast & HA & 0.8069 & 0.7694 & \textbf{0.9271} & 0.882 & - \\
\hline
\end{tabular}
}
\end{table}

Table \ref{tab:Table_10} and \ref{tab:Table_11} provide comparative results of the approaches of imputing null values with 0 and 2, respectively. As can be seen, when the null values are imputed with 0, ECRGNN performs a bit poorly as compared to the state-of-the-art models due to the dataset being biased more towards 0. However, this factor is mitigated, and ECRGNN starts performing better with a separate classification category for the null values as presented in \ref{tab:Table_11}. Since the Sider \cite{Cit_33} dataset does not have any missing values, ECRGNN outperforms the SOTA models in the corresponding dataset.

\subsection{Limitations and Future Work}
\par A detailed study of the novel ECRGNN, as well as the comparison, has been presented in this work. However, the study is not all-encompassing, and some limitations do exist. Though the GNN model has been fine-tuned, as seen from the regression results, the VAE architecture has not been hyper-parameter optimized for each corresponding dataset. Proper tuning of the hyperparameters is required for the VAE+GNN model to accurately measure the extent of ECRGNN's capabilities. Proper ablation studies with various types of VAEs are also required.

\par The GRU-based ECRGNN model, as the highlight, improves the long-range molecular dependencies of the model. In this study, experimentation was done with only GRU-based mechanics. Although the GRU-based model provides improved performance, experimentation should also be done with other types of Recurrent Neural Networks. This has been reserved as a part of a future study. Apart from this, feature engineering and reduction methods to reduce the label space complexity are also among the directions intended to be pursued in a future study.

\section*{Conclusions}
This paper presents a new Residual Graph model, the ECRGNN, to overcome the limitations of the ECC and GNN models. An end-to-end learning and analysis of the model has been presented on various SMILE-encoded molecular datasets for both classification and regression tasks. ECRGNN provides adequate improvement from the current SOTA, especially in cases of the Lipophilicity and Boiling Point datasets. Apart from that, special focus has been placed on the context of multi-class, multi-label classification of various datasets like SIDER, ToxCast, and Tox21. Statistical analyses of used datasets and various imputation techniques have been compared, demonstrating that the addition of separate labels to the missing values helps improve the prediction results. A VAE with the ECRGNN as a backbone has been presented in this context, which has increased the accuracy of the association of labels with molecular feature embeddings. Ablation studies show the ECRGNN performs quite competitively in this setting as well. The relationship of the ECRGNN results with the underlying chemistry of the molecules has been shown through the GNNExplainer algorithm. ECRGNN is observed to adhere better to the notion of the underlying chemistry than the ECC model on which it is based. ECRGNN still lacks transparency and ablation studies from a variety of VAEs and other possible pipeline algorithms. A comprehensive study on the impact of residual connections, the number of layers, and its impact on property prediction will be undertaken as a future work.

\section*{Author contributions}

\textbf{Kanad Sen}: Conceptualization, Methodology, Validation, Investigation, Coding, Writing – original draft.\\
\textbf{Saksham Gupta}:  Conceptualization, Methodology, Validation, Investigation, Coding, Writing – original draft.\\
\textbf{Abhishek Raj}: Data Curation, Visualization, Validation, Investigation, Writing – original draft.\\
\textbf{Alankar Alankar}: Writing – review and editing, Validation, Supervision

\section*{Conflicts of interest}
The authors declare no conflict of interest.

\section*{Data availability}

All the datasets that have been used have been mentioned and are publicly available. The compiled and formatted datasets can be provided upon valid requests.
\begin{acknowledgement}

The authors thank Prof. Alankar Alankar for the guidance, support, and constructive feedback received during this project. The authors also thank IIT Bombay for providing us with the necessary computational resources.

\end{acknowledgement}

\bibliography{ECRGNN}

\end{document}